# Energy Efficient Fog based Healthcare Monitoring Infrastructure

Ida Syafiza M. Isa, Taisir E.H. El-Gorashi, Mohamed O.I. Musa, and Jaafar M.H. Elmirghani

**Abstract—Recent advances in mobile technologies and cloud computing services have inspired the development of cloud-based real-time health monitoring systems. However, the transfer of health-related data to the cloud contributes to the burden on the networking infrastructures, leading to high latency and increased power consumption. Fog computing is introduced to relieve this burden by bringing services to the users' proximity. This study proposes a new fog computing architecture for health monitoring applications based on a Gigabit Passive Optical Network (GPON) access network. An Energy-Efficient Fog Computing (EEFC) model is developed using Mixed Integer Linear Programming (MILP) to optimize the number and location of fog devices at the network edge to process and analyze the health data for energy-efficient fog computing. The performance of the EEFC model at low data rates and high data rates health applications is studied. The outcome of the study reveals that a total energy saving of 36% and 52% are attained via processing and analysis the health data at the fog in comparison to conventional processing and analysis at the central cloud for low data rate application and high data rate application, respectively. We also developed a real-time heuristic; Energy Optimized Fog Computing (EOFC) heuristic, with energy consumption performance approaching the EEFC model. Furthermore, we examined the energy efficiency improvements under different scenarios of devices idle power consumption and traffic volume.**

## I. INTRODUCTION

The recent increase in chronic diseases, the ageing population and the increasing costs of healthcare have led to the revolution of remote health monitoring in developed countries [1]. The advances in wireless body sensors and mobile technologies have motivated the development of mobile-based health monitoring systems (m-Health) that provide real-time feedback to the patients pertaining to their health conditions and alerts in health-threatening conditions. Additionally, the rapid growth in cloud computing has enabled the development of mobile cloud computing (MCC) applications that offer higher processing and better storage capabilities to health data, aside from increasing the accuracy of the diagnosis. Furthermore, healthcare systems can be enhanced by using machine learning methods to perform early-detection and prediction of diseases [2]. Several cloud-based remote health monitoring systems have been developed. For instance, an m-Health monitoring system based on cloud computing platform (Could-MHMS), is designed for pervasive health monitoring among elderly patients [3]. The system leverages the high processing and storage of the cloud computing platform, whereby feasible and flexible personalized public healthcare services can be provided. In line with that, a real-time cloud-based system for users with mobile devices or web browsers was proposed in [4]. The proposed system has been proven to be functional, accurate, and efficient in both monitoring and analysing health data. Nonetheless, the transfer of large health-related data from a large number of patients to the cloud contributes to increasing the congestion in networking infrastructure which leads to high latency and hence violations of Quality-of-Service (QoS) metrics [5]. This also increases the occurrences of bit errors where the impact of a single error can cause inaccurate treatment decisions, which can be critical especially for emergency cases [6]. Furthermore, the large volumes of transferred data can increases the energy consumption within the network as the data has to travel multiple hops over the network to reach the cloud [7].

One effective way to address the limitations of cloud-based systems is to provision the service closer to users [8]. A new paradigm, referred to as fog computing, has been introduced [9] to extend cloud services by initiating an intermediate layer between end users and the cloud where processing and storage resources equipped with communication capabilities are available [10]. Connections to the cloud are also possible by the fog server to leverage the rich functionalities and applications of the cloud. Furthermore, fog nodes can be distributed at the network edge with dense geographical coverage and mobility support. Therefore, fog computing can deliver QoS metrics for healthcare monitoring systems for patients due to reduced network traffic and shorter network travel reducing the energy consumption in cloud networking infrastructures [11]–[19].

Several recent studies have applied fog computing to develop efficient health monitoring systems. For instance, a monitoring system in [6] employed the concept of fog computing at smart gateways to efficiently process health data, particularly electrocardiogram (ECG) signals. The ECG empirical results for feature extraction using the proposed system displayed 90% bandwidth efficiency and low latency real-time response. Additionally, the authors of [2] claimed that both continuous monitoring and real-time monitoring may be dysfunctional with the present limited processing IoT-based systems. Hence, fog computing was embedded into the monitoring system and the results exhibited reduced response time and increased system reliability in intermittent Internet connections. In [20], a prototype of a smart e-health gateway (i.e. the fog computing device) has been implemented to reduce the burden on the sensor node processing resources and the cloud by performing high-level services such as real-time data processing, local storage, and embedded data mining in the fog. The performance of the gateway is evaluated in terms of energy efficiency of the sensor nodes, scalability, mobility, and reliability. Furthermore, in [21] a real-time event-triggered health monitoring system for smart homes is implemented using a Bayesian Belief Network (BBN) to classify the events at the fog layer.

To the best of our knowledge, no studies have considered the essential aspect of the energy consumption in the access networks, transport networks, fog and cloud for healthcare applications, which are growing in number and importance. In this work, we develop a framework for energy efficient fog based real-time health monitoring systems. This framework is based on our previous research efforts on developing energy efficient architectures for cloud data centres and core networks [11], [16], [22]–[24]. We considered different techniques and technologies such as virtualization [12], [25], [26], network architecture design and optimization [15], [27]–[31], optimizing content distribution [13], [14], [32], progressive big data processing [19], [33]–[35], network coding [18], [36] and using renewable energy [17]. In [37], we showed that energy consumption is minimized when performing the processing and analysis of health data at the network edge compared to processing at the central cloud. In this paper, we further explore the energy efficiency potential of fog based health monitoring systems. The contribution of this paper can be summarized as follows:

i. Development of a detailed mathematical model using MILP to optimize the placement of processing servers (PSs) at the access network so that the total energy consumption of the health monitoring application is minimized.
ii. Development of a heuristic algorithm for real-time implementation of the energy efficient fog based health system.
iii. Evaluation of healthcare applications at different data rates: ECG monitoring (low data rate) and video based fall monitoring (high data rate) in a realistic case study.
iv. Study of the impact of networking equipment and servers idle power and the increasing traffic on the energy efficiency of the proposed healthcare monitoring system.

The remainder of the paper is organized as follows: Section II presents the proposed fog-based health monitoring system. Next, Section III presents the MILP model for energy efficient Fog-based health monitoring system. The parameters selection considered in this work are elaborated in Section IV. The performance of the proposed approaches for low data rate health monitoring application and the development of the heuristic are presented in Section V. The performance evaluation of the proposed approaches for high data rate health monitoring applications is presented in Section VI. Finally, this paper is concluded in Section VII.

## II. The proposed Fog-based health monitoring system

This section presents the proposed fog-based health monitoring system. The system is composed of three modules; health data analysis and decision-making module, fog storage module and cloud storage module, as illustrated in Figure 1. The health data analysis and decision-making and the fog storage are embedded in the fog layer while the cloud storage is incorporated in the cloud layer.

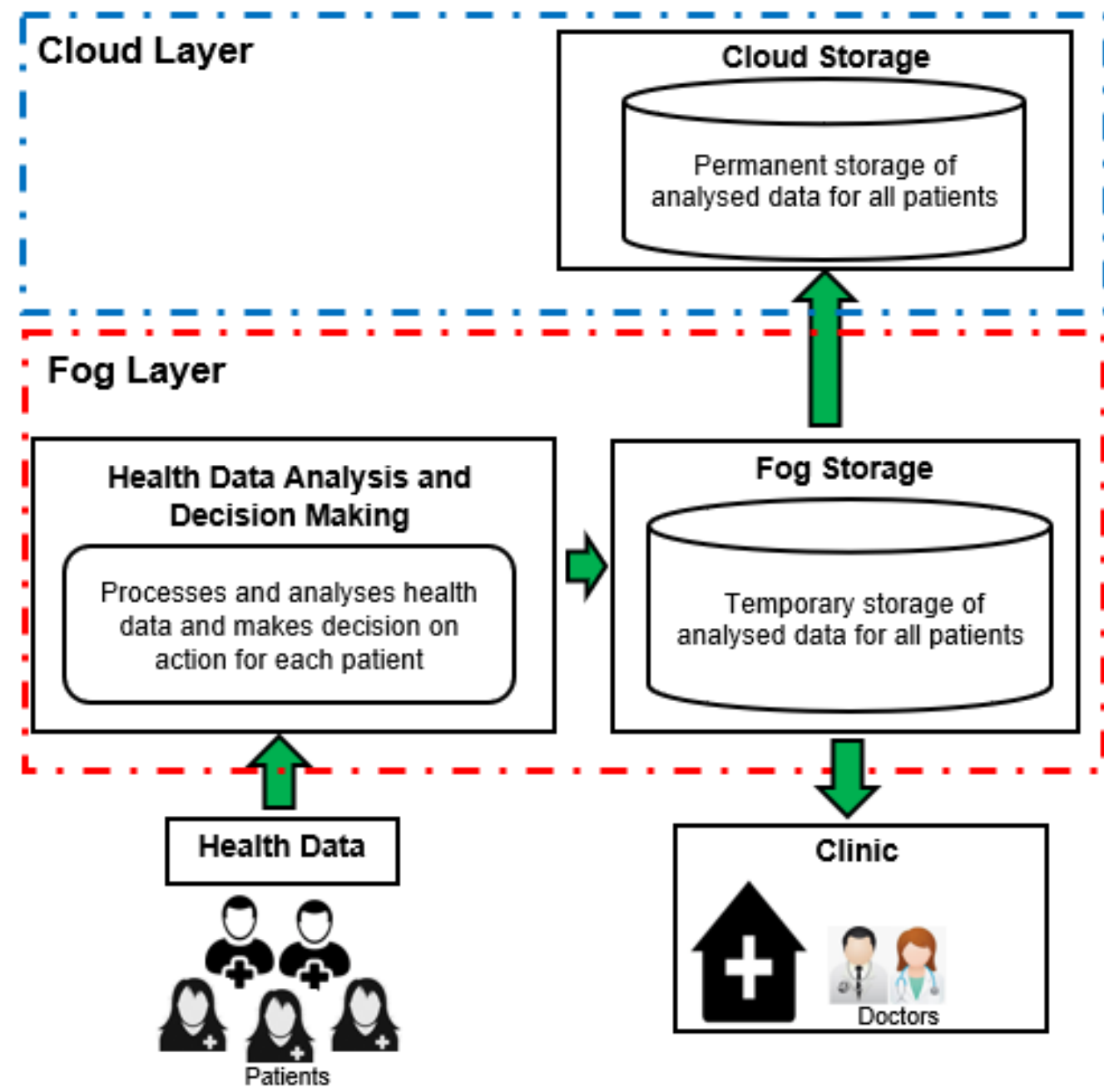

Figure 1: Architecture of the proposed system

The *Fog Storage module* is a temporary storage unit for the analyzed health data before being sent to the cloud for permanent storage. The *fog storage module* provides alternative access to health data for urgent demands. This module is also used to send the analyzed health data to the cloud storage and the clinic for permanent storage and feedback purposes, respectively. The *Health Data Analysis and Decision-making Module* performs three tasks. The first is aggregating the health data sent from multiple patients via wireless-connected devices. The second task is the processing and analyses of patients' health data and matching it with disease symptoms based on the extracted features. The final task is making decisions on the action taken against irregular physical data of the patients, such as informing the emergency medical service resources to act fast for patients who need aid. Nonetheless, in some cases, the doctors would re-diagnosis the results before making the final decision. The *Cloud Storage Module* permanently stores the analyzed results of patients for medical records purposes. This module offers accessibility for both patients and doctors, similar to that in the fog storage module. Figure 2 illustrates the system flow of the proposed fog based architecture (Figure 2(a)) and a cloud-based architecture (Figure 2(b)) where the raw health data is sent to the central cloud for processing and analysis, feedback is sent from the cloud to the patient/clinic and the analyzed data is permanently stored in the cloud. Note that the three tasks: the processing and analysis task, the feedback task and the storage task occur at different times.

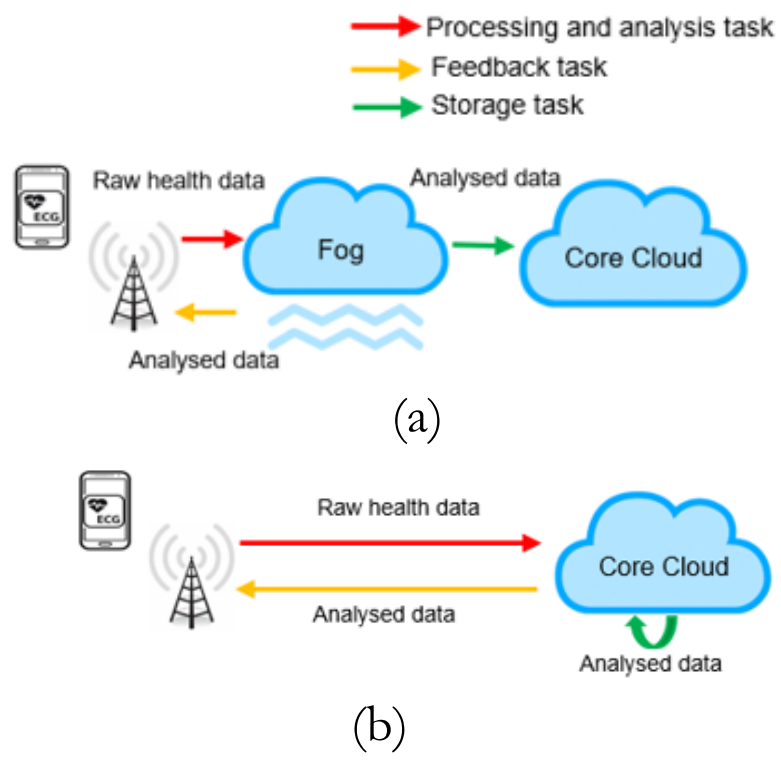

Figure 2: System flow of (a) proposed approach (FOA) (b) conventional approach (CA)

The fog based health monitoring system is to be embedded in a network architecture characterized by four layers, as portrayed in Figure 3:

- IoT Devices Layer: This is the bottom-most layer. It is comprized of IoT devices, mobile phones, iPads etc. with Machine-to-Machine (M2M) communication capabilities to connect to wireless body sensors to monitor patients and to send data to higher layers of the network.
- Access Layer: This is the access network layer where the fog processing resources reside. This layer aggregates data from the IoT layer via gateways such as an LTE-M base station, Wi-Fi access point, etc. A GPON network [38] is considered to connect the gateways to the higher layers of the architecture. Fog processing resources serving the health monitoring application can be deployed at ONUs and/or the OLT of the GPON which are equipped with an internal switch. Placing the processing servers (PSs) at ONUs, which is closer to the users, decreases the energy consumption of the networking equipment, however, it will increase the required number of PSs. On the other hand, deploying PSs at the OLT reduces the number of PSs required as it is accessible by all users via minimum number of network hops. This, however, will increase the energy consumption of the networking equipment. Also, an additional Ethernet switch is used at the ONUs and OLT in scenarios where more than one PSs are connected to the same node.

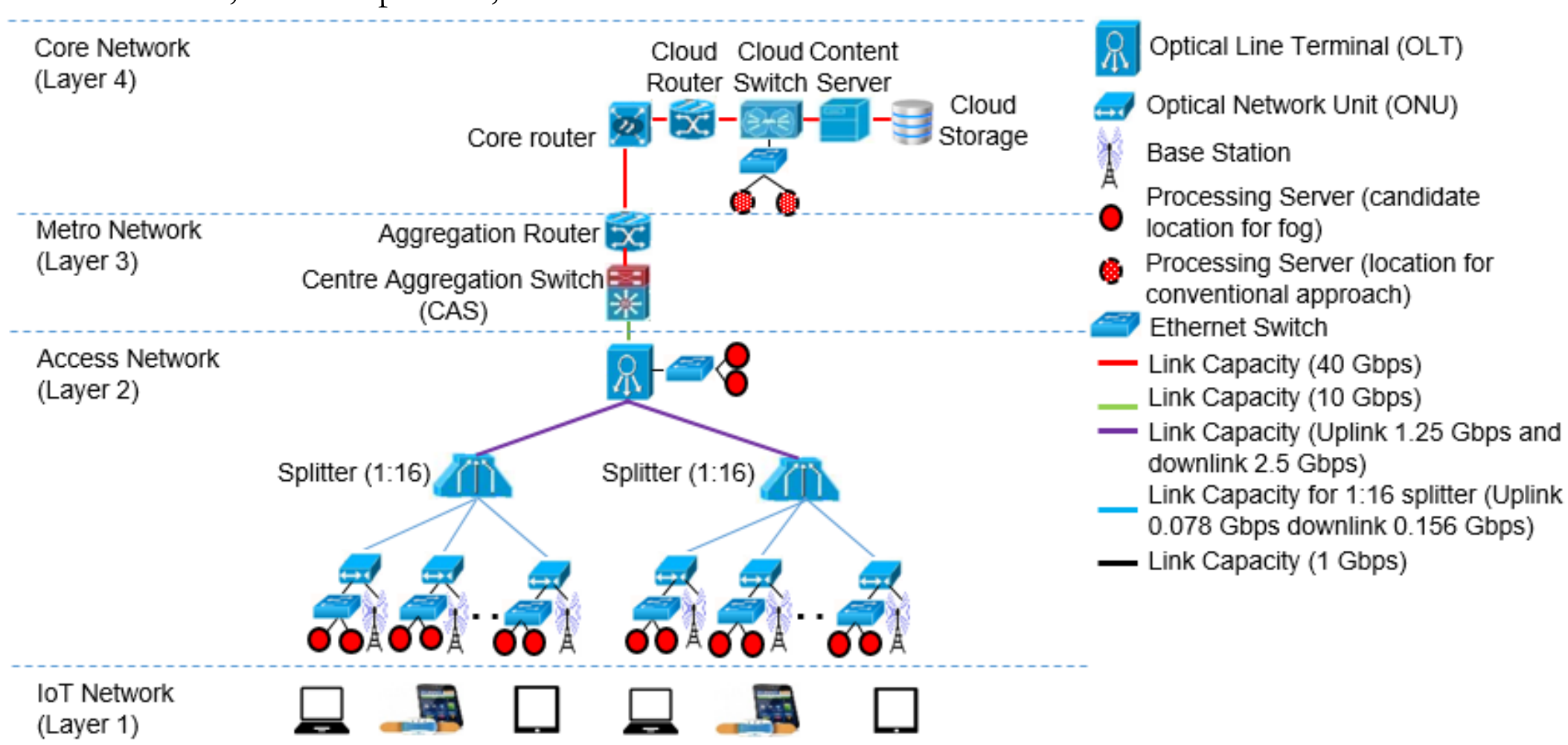

Figure 3: GPON architecture in the Fog Network

- Metro Layer: In this layer, a centre aggregation switch (CAS) and aggregation router are used to aggregate traffic from access networks and forward traffic to the upper layer. The CAS aggregates and fast-forwards data between the BSs in the access network. The aggregation router serves as a gateway to connect the access network to the core network.

  Core Layer: This is the upper-most layer in the architecture based on an IP over WDM architecture. The most power-consuming devices in an IP over WDM node is the IP router. Cloud data centres are connected to the core network nodes. Inside the data centre routers and switches, are used to connect content servers and cloud storages.

## III. MILP MODEL FOR ENERGY-EFFICIENT FOG COMPUTING HEALTH MONITORING SYSTEM USING LTE-M (EEFC)

This section presents the MILP model developed to minimize the networking and processing energy consumption of the fog optimized approach (FOA) by optimizing the location of PSs in the access network. The networking energy consumption includes the energy consumed by networking devices at all layers while the processing energy consumption refers to the energy consumed by the PSs. The architecture considered uses LTE-M base stations (BSs) to aggregate traffic from the IoT layer. Before introducing the model, we define the sets, parameters and variables used as follows:

Table 1: The sets, parameters and variables used in MILP

| Sets | |
|---|---|
| $CL$ | Set of clinics |
| $BS$ | Set of BSs |
| $ONU$ | Set of ONUs |
| $OLT$ | Set of OLTs |
| $CAS$ | Set of centre aggregation switches |
| $AR$ | Set of aggregation routers |
| $CR$ | Set of core routers |
| $CLR$ | Set of cloud routers |
| $CLS$ | Set of cloud switches |
| $CS$ | Set of content servers |
| $CST$ | Cloud storage |
| $N_m$ | Set of neighbouring nodes of node $m$ in the network |
| $N$ | Set of nodes ($N \in CL \cup BS \cup ONU \cup OLT \cup CAS \cup AR \cup CR \cup CLR \cup CLS \cup CS \cup CST$) |
| $FN$ | Set of candidate locations to deploy PS (fog) ($FN \in ONU \cup OLT$) |
| Parameters | |
| $s$ and $d$ | Denote source node $s$ and destination node $d$ of traffic between a node pair |
| $i$ and $j$ | Denote end nodes of a physical link in the network, $i, j \in N$ |
| $Pt_s$ | Number of patients in clinic $s$ |
| $IBS$ | Idle power consumption of a base station (W) |
| $PBS$ | Power per physical resource block (PRB) of a base station (W/PRB) |
| R | Maximum number of PRBs in a base station dedicated for healthcare applications |
| $PONU$ | Maximum power consumption of an ONU (W) |
| $IONU$ | Idle power consumption of an ONU (W) |
| $CONU$ | Maximum capacity of an ONU (bps) |
| $PES$ | Maximum power consumption of an Ethernet switch (W) |
| $IES$ | Idle power consumption of an Ethernet switch (W) |
| $CES$ | Maximum capacity of an Ethernet switch (bps) |
| $POLT$ | Maximum power consumption of an OLT (W) |
| $IOLT$ | Idle power consumption of an OLT (W) |
| $COLT$ | Maximum capacity of an OLT (bps) |
| $PCAS$ | Maximum power consumption of a centre aggregation switch (W) |
| $ICAS$ | Idle power consumption of a centre aggregation switch (W) |
| $CCAS$ | Maximum capacity of a centre aggregation switch (bps) |
| $PAR$ | Maximum power consumption of an aggregation router (W) |
| $IAR$ | Idle power consumption of an aggregation router (W) |
| $CAR$ | Maximum capacity of an aggregation router (bps) |
| $PCR$ | Maximum power consumption of a core router (W) |
| $ICR$ | Idle power consumption of a core router (W) |
| $CCR$ | Maximum capacity of a core router (W) |
| $PCLR$ | Maximum power consumption of a cloud router (W) |
| $ICLR$ | Idle power consumption of a cloud router (W) |
| $CCLR$ | Maximum capacity of a cloud router (bps) |
| $PCLS$ | Maximum power consumption of a cloud switch (W) |
| $ICLS$ | Idle power consumption of a cloud switch (W) |
| $CCLS$ | Maximum capacity of a cloud switch (bps) |
| $PCS$ | Maximum power consumption of a content server (W) |
| $ICS$ | Idle power consumption of a content server (W) |
| $CCS$ | Maximum capacity of a content server (bps) |
| $PCST$ | Maximum power consumption of cloud storage (W) |
| $ICST$ | Idle power consumption of cloud storage (W) |
| $CCST$ | Maximum capacity of cloud storage (bits) |
| $PPS$ | Maximum power consumption of a PS (W) |
| $IPS$ | Idle power consumption of a PS (W) |
| $\Omega max$ | Maximum number of patients per PS |
| $\Lambda max$ | Maximum storage capacity of PS (bits) |
| $\delta a$ | Data rate per patient to send raw health data from clinic to PS (bps) |
| $\tau a$ | Transmission time per patient to send raw health data from clinic to PS (s) |
| $Ra$ | Physical resource block per patient to send raw health data from clinic to PS |
| $\alpha$ | Size of analyzed health data per patient (bit) |
| $\delta b$ | Data rate per patient to send analyzed health data from PS to clinic (bps) |
| $\tau b$ | Transmission time per patient to send analyzed health data from PS to clinic (s) |
| $Rb$ | Physical resource block per patient to send analyzed health data from PS to clinic |
| $\delta c$ | Data rate per patient to send analyzed health data from PS to cloud storage (bps) |
| $\tau c$ | Transmission time per patient to send analyzed health data from PS to cloud storage (s) |
| $\delta_{sd}$ | $\delta_{sd} = 1$ to send the storage traffic from PSs located at node $s$, to the cloud storage node $d$, $s \in FN, d \in CST$ |
| $x$ | Fraction of idle power consumption of networking equipment contributed by the healthcare application under consideration |
| $\lambda_{ij}$ | The capacity of link $ij$ dedicated for the healthcare application under consideration (bps) |
| $\eta$ | Power usage effectiveness (PUE) of the access network, metro network and IP over WDM network |
| $c$ | Power usage effectiveness (PUE) of the fog (PS) and cloud equipment |
| $M$ | A large enough number |
| Variables | |
| $P_{sd}$ | Raw health data traffic from source node $s$ to destination node $d$ (bps), $s \in CL, d \in FN$ |
| $P_{ij}^{sd}$ | Raw health data traffic from source node $s$ to destination node $d$ that traverses the link between nodes $i$ and $j$ (bps), $s \in CL, d \in FN,\ i, j \in N$ |
| $P_i$ | Total raw health data traffic that traverses node $i$ (bps), $i \in N$ |
| $F_{sd}$ | Analyzed health data feedback traffic from source node $s$ to destination node $d$ (bps), $s \in FN, d \in CL$ |
| $F_{ij}^{sd}$ | Analyzed health data feedback traffic from source node $s$ to destination node $d$ that traverses the link between nodes $i$ and $j$ (bps), $s \in FN, d \in CL, i, j \in N$ |
| $F_i$ | Total analyzed health data feedback traffic that traverses node $i$ (bps), $i \in N$ |
| $S_{sd}$ | Analyzed health data storage traffic from source node $s$ to destination node $d$ (bps), $s \in FN, d \in CST$ |
| $S_{ij}^{sd}$ | Analyzed health data storage traffic from source node $s$ to destination node $d$ that traverses the link between nodes $i$ and $j$ (bps), $s \in FN, d \in CST, i, j \in N$ |
| $S_i$ | Total analyzed health data storage traffic that traverses node $i$ (bps), $i \in N$ |
| $\omega_{sd}$ | Number of patients from clinic $s$ served by PS located at node $d$ |
| $Pa_{ij}$ | Number of patients in clinic $i$ served by BS $j$ to send raw health data traffic (integer) |
| $Pb_{ij}$ | Number of patients in clinic $i$ served by BS $j$ to receive analyzed health data feedback traffic (integer) |
| $\beta a_j$ | Number of PRBs used in BS $j$ to serve raw health data traffic (integer) |
| $\beta b_i$ | Number of PRBs used in BS $i$ to serve analyzed health data feedback traffic (integer) |
| $Y_d$ | $Y_d = 1$, if one or more PSs are placed at node $d$, otherwise $Y_d = 0$, $d \in FN$ |
| $\phi_d$ | Number of PSs placed at node $d$, $d \in FN$ |
| $\tau p_d$ | Processing and analysis time of PS (seconds) at node $d$, $d \in FN$ |
| $\zeta a_j$ | $\zeta a_j = 1$, if raw health data traffic traverses node $j$, otherwise $\zeta a_j = 0, j \in N$ |
| $\zeta b_i$ | $\zeta b_i = 1$, if analyzed health data feedback traffic traverses node $i$, otherwise $\zeta b_i = 0, i \in N$ |
| $\theta c_i$ | $\theta c_i = 1$, if analyzed health data storage traffic traverses node $i$ where node $i$ is the source of a link, otherwise $\theta c_i = 0, i \in N$ |
| $\vartheta c_j$ | $\vartheta c_j = 1$, if analyzed health data storage traffic traverses node $j$ where $j$ is the end of a link, otherwise $\vartheta c_j = 0, j \in N$ |
| $\zeta c_i$ | $\zeta c_i = 1$, if the analyzed health data storage traffic traverses node $i$ where $\zeta c_i = \theta c_i \; OR \; \vartheta c_i$, otherwise \ $\zeta c_i = 0, i \in N$ |
| $\nu_i$ | $\nu_i$ is a dummy variable that takes value of $\theta c_i \oplus \vartheta c_i$, where $\oplus$ is an XOR operation, $i \in N$ |
| $EAN$ | Energy consumption of access network |
| $ETBS$ | Total energy consumption of base stations |
| $EBSP$ | Energy consumption of base stations required to relay raw health data traffic |
| $EBSF$ | Energy consumption of base stations required to relay analyzed health data feedback traffic |
| $ETONU$ | Total energy consumption of ONUs |
| $EONUP$ | Energy consumption of ONUs required to relay raw health data traffic |
| $EONUF$ | Energy consumption of ONUs required to relay analyzed health data feedback traffic |
| $EONUS$ | Energy consumption of ONUs required to relay analyzed health data storage traffic |
| $ETES$ | Total energy consumption of Ethernet switches |
| $EESP$ | Energy consumption of Ethernet switches required to relay raw health data traffic |
| $EESF$ | Energy consumption of Ethernet switches required to relay analyzed health data feedback traffic |
| $EESS$ | Energy consumption of Ethernet switches required to relay analyzed health data storage traffic |
| $ETOLT$ | Total energy consumption of OLTs |
| $EOLTP$ | Energy consumption of OLTs required to relay raw health data traffic |
| $EOLTF$ | Energy consumption of OLTs required to relay analyzed health data feedback traffic |
| $EOLTS$ | Energy consumption of OLTs required to relay analyzed health data storage traffic |
| $EMN$ | Energy consumption of metro network |
| $ECASS$ | Energy consumption of centre aggregation switches required to relay analyzed health data storage traffic |
| $EARS$ | Energy consumption of aggregation routers required to relay analyzed health data storage traffic |
| $ECN$ | Energy consumption of core network |
| $ECRS$ | Energy consumption of core routers required to relay analyzed health data storage traffic |
| $ECL$ | Energy consumption of cloud |
| $ECLRS$ | Energy consumption of cloud routers required to relay analyzed health data storage traffic |
| $ECLSS$ | Energy consumption of cloud switches required to relay analyzed health data storage traffic |
| $ECSS$ | Energy consumption of content servers required to relay analyzed health data storage traffic |
| $ECSTS$ | Energy consumption of cloud storage required to store the analyzed health data storage traffic |
| $EFN$ | Energy consumption of fog nodes |
| $EPS$ | Energy consumption of PSs |

We start by defining the energy consumption of the network, including access, metro and core networks, and PSs at fog nodes and cloud.

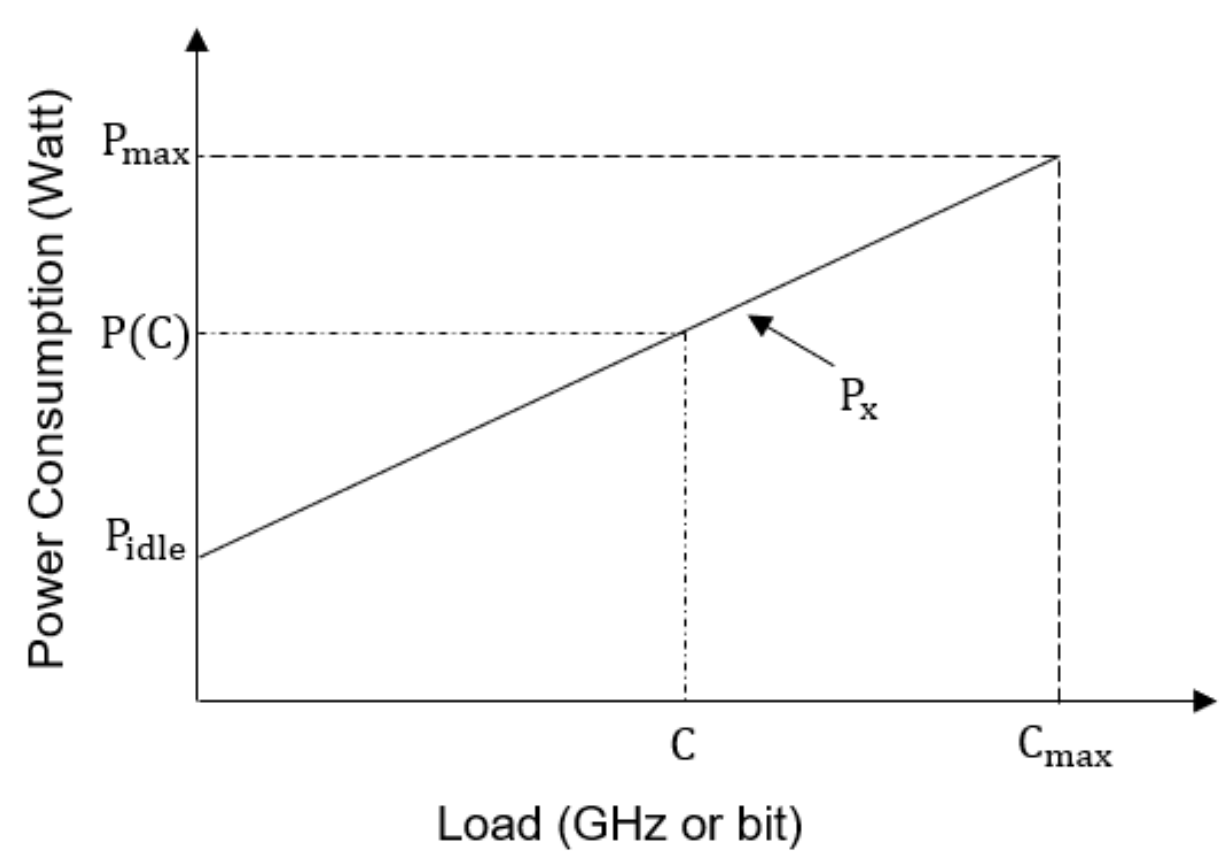

(a)

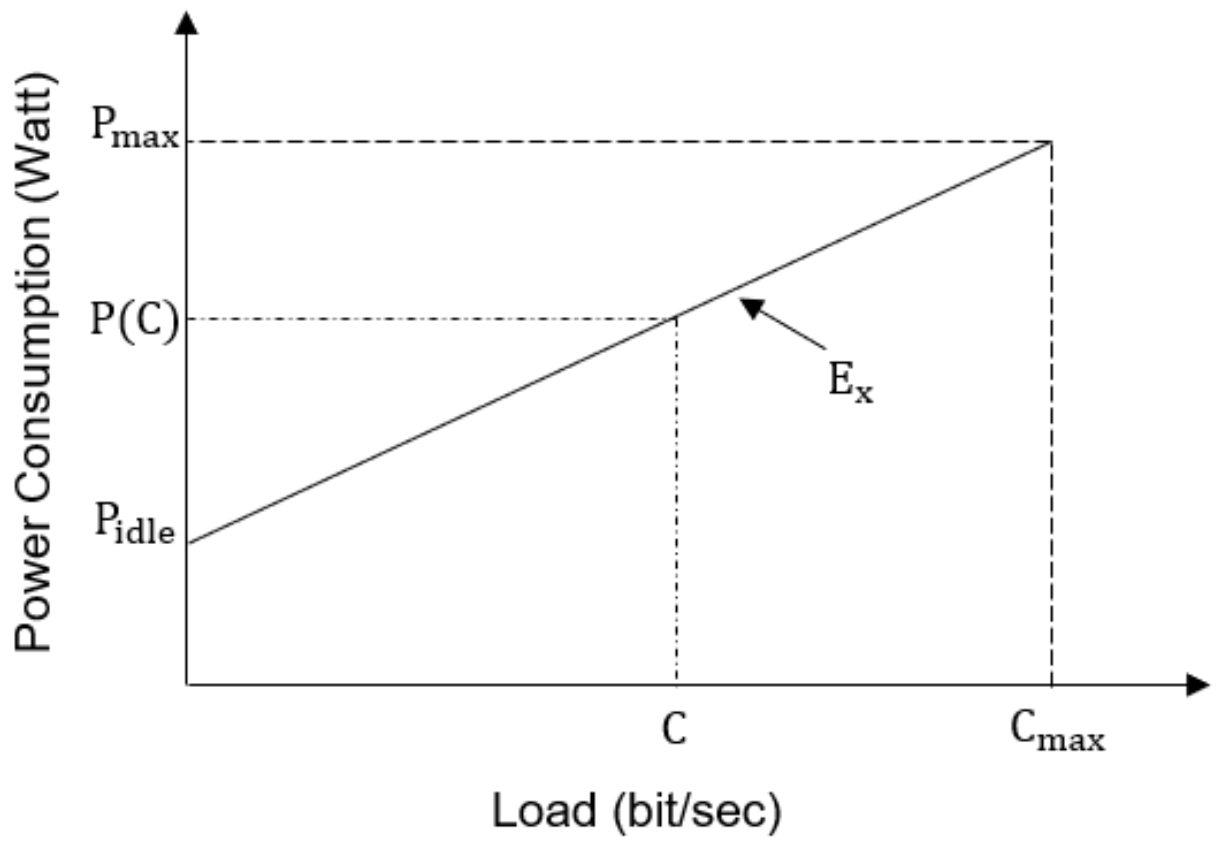

(b)

Figure 4: Power consumption profile for (a) BS, PS and cloud storage (b) ONU, OLT, Ethernet switch, centre aggregation switch, aggregation router, core router, cloud router, cloud switch and content server .

Figure 4-(a) illustrates the power profile for PS, BS and cloud storage while Figure 4-(b) illustrates the power profile for the other devices. As for BS, PS, and cloud storage; the power consumption is composed of a fixed idle power and the load-dependent power given as:

$$P(C) = P_{idle} + C\,\frac{P_{max} - P_{idle}}{C_{max}} = P_{idle} + C\,P_x \tag{1}$$

where $P_{idle}$ denotes idle power consumption, and the graph slope ($P_x$) represents power consumption per physical resource block (PRB) for BS, power per GHz for PS and power per Gbit for cloud storage. $C$ refers to the load in RB, GHz and Gbits for BS, PS and cloud storage, respectively.

Equation (2) gives the power consumption of other networking devices including ONU, OLT, Ethernet switch, centre aggregation switch, aggregation router, core router, cloud router, cloud switch and content server:

$$P(C) = P_{idle} + C\,\frac{P_{max} - P_{idle}}{C_{max}} = P_{idle} + C\,E_b \tag{2}$$

where the slope of the graph ($E_b$) reflects the increase in energy per bit. Besides, $C$ denotes the offered load in bits per second.

In the following, we show the energy consumption of the network layers, cloud and fog. Note that the energy consumption at the IoT devices is not considered.

*a)* Energy consumption of access network, $EAN$:

The energy consumption of access network, $EAN$ is composed of energy of LTE BSs, ONUs and OLTs and is as given in (3):

$$EAN = (ETBS + EOTNU + ETOLT)\,\eta \tag{3}$$

The energy consumption of BSs, $ETBS,$ is calculated as follows:

$$ETBS = EBSP + EBSF \tag{4}$$

where

$$EBSP = \sum_{i\in BS} (IBS\ x\ \zeta a_i + PBS\ \beta a_i)\ \tau a \tag{5}$$

$$EBSF = \sum_{i\in BS} (IBS\ x\ \zeta b_i + PBS\ \beta b_i)\ \tau b \tag{6}$$

The energy consumed by LTE base stations is composed of the energy consumed to relay raw health data and the energy consumed to transmit feedback traffic which is a function of the number of PRBs and the time the BS is used to send traffic as given in equations (5) and (6), respectively. Note that $x$ refers to the fraction of idle power contributed by the healthcare application.

The energy consumption of ONUs, $ETONU$, is given as:

$$ETONU = EONUP + EONUF + EONUS \tag{7}$$

where

$$EONUP = \sum_{i\in ONU} \left(IONU\ x\ \zeta a_i + P_i\,\frac{(PONU - IONU)}{CONU}\right) \tau a \tag{8}$$

$$EONUF = \sum_{i\in ONU} \left(IONU\ x\ \zeta b_i + F_i\,\frac{(PONU - IONU)}{CONU}\right) \tau b \tag{9}$$

$$EONUS = \sum_{i\in ONU} \left(IONU\ x\ \zeta c_i + S_i\,\frac{(PONU - IONU)}{CONU}\right) \tau c \tag{10}$$

The energy consumed by the ONUs is composed of the energy consumed to relay the raw health data traffic, analyzed health data feedback traffic and analyzed health data storage traffic as detailed in equation (8)-(10), respectively.

The energy consumption of the OLT, $ETOLT$, is given as:

$$ETOLT = EOLTP + EOLTF + EOLTS \tag{11}$$

where

$$EOLTP = \sum_{i\in OLT} \left(IOLT\ x\ \zeta a_i + P_i\,\frac{(POLT - IOLT)}{COLT}\right) \tau a \tag{12}$$

$$EOLTF = \sum_{i\in OLT} \left(IOLT\ x\ \zeta b_i + F_i\,\frac{(POLT - IOLT)}{COLT}\right) \tau b \tag{13}$$

$$EOLTS = \sum_{i\in OLT} \left(IOLT\ x\ \zeta c_i + S_i\,\frac{(POLT - IOLT)}{COLT}\right) \tau c \tag{14}$$

The energy consumption of the OLT is composed of the energy consumed to relay raw health data traffic, analyzed health data feedback traffic, and analyzed health data storage traffic as detailed in equations (12)-(14), respectively.

*b)* Energy consumption of metro network, $EMN$

The energy consumption of metro network $EMN$ is composed of energy consumption of the central aggregation switch and aggregation router. Note that these devices are only used to relay the analyzed health data storage traffic as the candidate locations of PSs is at the access layer. The energy consumption of metro network, $EMN$, is as given follows:

$$EMN = (ECASS + EARS)\,\eta \tag{15}$$

where

$$ECASS = \sum_{i\in CAS} \left(ICAS\ x\ \zeta c_i + S_i\,\frac{(PCAS - ICAS)}{CCAS}\right) \tau c \tag{16}$$

$$EARS = \sum_{i\in AR} \left(IAR\ x\ \zeta c_i + S_i\,\frac{(PAR - IAR)}{CAR}\right) \tau c \tag{17}$$

*c)* Total energy consumption of core network, $ECN$

The energy consumption of core network, $ECN$ is composed of energy consumption of core routers to relay the analyzed health data storage traffic as given in Equation (18):

$$ECN = ECRS\,\eta \tag{18}$$

where

$$ECRS = \sum_{i \in CR} \left( ICR\ x\ \zeta c_i + S_i\ \frac{(PCR - ICR)}{CCR} \right) \tau c \tag{19}$$

*d)* Energy consumption of cloud, $ECL$

The energy consumption of cloud, $ECL$, is composed of energy of cloud routers, cloud switches, content servers and cloud storage. Note that cloud storage is used to perform the storage task while other devices are used to relay the analyzed health data storage traffic. The energy consumption of the cloud is given in Equation (20):

$$ECL = (ECLRS + ECLSS + ECSS + ECSTS)\ c \tag{20}$$

where

$$ECLRS = \sum_{i \in CLR} \left( ICLR\ x\ \zeta c_i + S_i\ \frac{(PCLR - ICLR)}{CCLR} \right) \tau c \tag{21}$$

$$ECLSS = 2 \sum_{i \in CLS} \left( ICLS\ x\ \zeta c_i + S_i\ \frac{(PCLS - ICLS)}{CCLS} \right) \tau c \tag{22}$$

$$ECSS = \sum_{i \in CS} \left( ICS\ x\ \zeta c_i + S_i\ \frac{(PCS - ICS)}{CCS} \right) \tau c \tag{23}$$

$$ECSTS = 2 \sum_{i \in CST} \left( ICST\ x\ \zeta c_i + S_i\ \tau c\ \frac{(PCST - ICST)}{CCST} \right) \tau c \tag{24}$$

Note that the energy consumption of the cloud switches and the cloud storage is multiplied by '2' for equipment redundancy purposes [32].

*e)* Energy consumption of fog nodes, $EFN$:

The energy consumed by the fog, $EFN$, reflects the energy consumed by PS, $EPS$, as given below:

$$EFN = EPS\ c + ETES\ \eta \tag{25}$$

where

$$EPS = \sum_{d \in FN} (IPS\ \phi_d\ (\tau a + \tau b + \tau c) + PPS\ \tau p_d) \tag{26}$$

$$ETES = EESP + EESF + EESS \tag{27}$$

$$EESP = \sum_{i \in FN} \left( IES\ x\ Y_i + P_i\ \ \frac{(PES - IES)}{CES} \right) \tau a \tag{28}$$

$$EESF = \sum_{i \in FN} \left( IES\ x\ Y_i + F_i\ \ \frac{(PES - IES)}{CES} \right) \tau b \tag{29}$$

$$EESS = \sum_{i \in FN} \left( IES\ x\ Y_i + S_i\ \ \frac{(PES - IES)}{CES} \right) \tau c \tag{30}$$

The idle energy consumption of the PSs is calculated considering the following: the time to receive raw health data from clinic, $\tau a$, the time to transmit the analyzed health data to clinics, $\tau b$, as well as the time to transmit the analyzed health data to cloud storage, $\tau c$. Note that we assume the PS works at full utilization to process the raw health data. The proportional energy consumption of processing and analysis for the PS is determined considering the time to perform the processing and analysis, $\tau p_d$. The energy consumption of Ethernet switches, $ETES$ is calculated considering the energy consumed to serve the raw health data traffic, analyzed health data feedback traffic and analyzed health data storage traffic as shown in Equations (28)-(30), respectively. Note that the energy consumed by the Ethernet switches is considered for a scenario where more than one PS can be connected to the ONU and OLT ($\phi_i$ is a variable). Also, note that the energy of the Ethernet switches is consumed if the utilized PSs are connected to it ($Y_i = 1$).

The model is defined as follows:
Objective:
Minimize the total energy consumption of networking and processing given as:

$$EAN + EMN + ECN + ECL + EFS \tag{31}$$

Subject to:

*1)* Allocating PSs to patients:

$$\omega_{sd} \leq Pt_s\ Y_d \quad ; \ \forall s \in CL, \forall d \in FN \tag{32}$$

Constraint (32) is used to allocate a fog node where one or more PSs are placed to serve patients of a clinic $s$. Note that patients of a clinic can by served by different fog nodes.

$$\sum_{d \in FN} \omega_{sd} = Pt_s \quad ; \ \forall s \in CL \tag{33}$$

Constraint (33) ascertains that each patient is served by a fog node.

*2)* Traffic from clinics to fog node.

$$P_{sd} = \omega_{sd}\ \delta a \quad ; \ s \in CL, d \in FN \tag{34}$$

Constraint (34) calculates the raw health data traffic from a clinic to a fog node based on the allocation of fog nodes to patients of the clinic as well as the uplink data rate provisioned to each patient.

*3)* Traffic from fog nodes to clinics.

$$F_{sd} = \omega_{ds}\ \delta b \quad ; \ \forall s \in FN, d \in CL \tag{35}$$

Constraint (35) calculates the analyzed health data feedback traffic from a fog node to a clinic based on the allocation of fog nodes to patients of the clinic as well as the downlink data rate provisioned to each patient.

*4)* Traffic from fog nodes to cloud storage.

$$S_{sd} = \sum_{i \in CL} \omega_{is}\ \delta c\ \delta_{sd} \quad ; \ \forall s \in FN, d \in CST \tag{36}$$

Constraint (36) calculates the analyzed health data storage traffic from a fog node to cloud storage based on the total number of patients served by the fog node and the data rate provisioned for each patient to send analyzed health data from PS to cloud storage. Note that in this work we only utilize one cloud storage, hence, $\delta_{sd}$=1.

*5)* Flow conservation in the network.

$$\sum_{j\in Nm[i]:i\neq j} P_{ij}^{sd} - \sum_{j\in Nm[i]:i\neq j} P_{ji}^{sd} = \begin{cases} P_{sd} \; if \; i = s \\ -P_{sd} \; if \; i = d \\ 0 \; otherwise \end{cases} \quad (37)$$
$$s \in CL, d \in FN, i \in N$$

$$\sum_{j\in Nm[i]:i\neq j} F_{ij}^{sd} - \sum_{j\in Nm[i]:i\neq j} F_{ji}^{sd} = \begin{cases} F_{sd} \; if \; i = s \\ -F_{sd} \; if \; i = d \\ 0 \; otherwise \end{cases} \quad (38)$$
$$s \in FN, d \in CL, i \in N$$

$$\sum_{j\in Nm[i]:i\neq j} S_{ij}^{sd} - \sum_{j\in Nm[i]:i\neq j} S_{ji}^{sd} = \begin{cases} S_{sd} \; if \; i = s \\ -S_{sd} \; if \; i = d \\ 0 \; otherwise \end{cases} \quad (39)$$
$$s \in FN, d \in CST, i \in N$$

Constraints (37)-(39) ensure that the total incoming traffic is equivalent to the total outgoing traffic for all nodes in the network, except for source and destination nodes for processing, feedback and storage traffic, respectively.

*6)* Total traffic traversing a node.

$$P_i = \left( \sum_{s\in CL} \sum_{d\in FN:s\neq d} \sum_{j\in Nm[i]:i\neq j} P_{ji}^{sd} \right) \quad ; \; \forall i \in N \quad (40)$$

$$F_i = \left( \sum_{s\in FN} \sum_{d\in CL:s\neq d} \sum_{j\in Nm[i]:i\neq j} F_{ij}^{sd} \right) \quad ; \; \forall i \in N \quad (41)$$

$$S_i = \left( \sum_{s\in FN} \sum_{d\in CST:s\neq d} \sum_{j\in Nm[i]:i\neq j} S_{ji}^{sd} + \sum_{d\in CST:i\neq d} S_{id} \right) \quad ; \; \forall i \in N \quad (42)$$

Constraints (40)-(42) calculate the total raw health data traffic, analyzed health data feedback traffic, and analyzed health data storage traffic that traverses node $i$, respectively.

*7)* Link capacity constraint.

$$\sum_{s\in CL} \sum_{d\in FN} P_{ij}^{sd} \leq \lambda_{ij} \quad ; \; \forall i \in N, \forall j \in Nm[i]: i \neq j \quad (43)$$

$$\sum_{s\in FN} \sum_{d\in CL} F_{ij}^{sd} \leq \lambda_{ij} \quad ; \; \forall i \in N, \forall j \in Nm[i]: i \neq j \quad (44)$$

$$\sum_{s\in FN} \sum_{d\in CST} S_{ij}^{sd} \leq \lambda_{ij} \quad ; \; \forall i \in N, \forall j \in Nm[i]: i \neq j \quad (45)$$

Constraints (43)-(45) ensure that the capacity of physical links used to send the total raw health data from clinics to fog nodes, the total analyzed health data from fog nodes to clinics for feedback, and the total analyzed health data from fog nodes to the cloud storage, respectively, does not exceed the maximum capacity of the links. Note that, as mentioned above, the three tasks occur at different times.

*8)* Node used to transmit the raw health data traffic from clinic to PS.

$$\sum_{s\in CL} \sum_{d\in FN} \sum_{i\in N:i\neq j} P_{ij}^{sd} \geq \zeta a_j \quad ; \; \forall j \in N \quad (46)$$

$$\sum_{s\in CL} \sum_{d\in FN} \sum_{i\in N:i\neq j} P_{ij}^{sd} \leq M \, \zeta a_j \quad ; \; \forall j \in N \quad (47)$$

Constraints (46) and (47) identify the nodes traversed by the raw health data traffic from clinics to fog nodes.

*9)* Node used to transmit the analyzed health data feedback traffic from PS to clinic

$$\sum_{s\in FN} \sum_{d\in CL} \sum_{j\in Nm[i]:i\neq j} F_{ij}^{sd} \geq \zeta b_i \quad ; \; \forall i \in N \quad (48)$$

$$\sum_{s\in FN} \sum_{d\in CL} \sum_{j\in Nm[i]:i\neq j} F_{ij}^{sd} \leq M \, \zeta b_i \quad ; \; \forall i \in N \quad (49)$$

Constraints (48) and (49) ensure $\zeta b_i = 1$ if the analyzed health data feedback traffic traverses node $i$ to send the analyzed data from PSs at node $s$ to clinics $d$, otherwise it is zero.

*10)* Node used to transmit the analyzed health data storage traffic from PS to cloud storage.

$$\sum_{s\in FN} \sum_{d\in CST} \sum_{j\in Nm[i[:i\neq j} S_{ij}^{sd} \geq \theta c_i \quad ; \; \forall i \in N \quad (50)$$

$$\sum_{s\in FN} \sum_{d\in CST} \sum_{j\in Nm[i]:i\neq j} S_{ij}^{sd} \leq M \, \theta c_i \quad ; \; \forall i \in N \quad (51)$$

$$\sum_{s\in FN} \sum_{d\in CST} \sum_{i\in Nm[j]:i\neq j} S_{ij}^{sd} \geq \vartheta c_j \quad ; \; \forall j \in N \quad (52)$$

$$\sum_{s\in FN} \sum_{d\in CST} \sum_{i\in Nm[j]:i\neq j} S_{ij}^{sd} \leq M \, \vartheta c_j \quad ; \; \forall j \in N \quad (53)$$

$$\theta c_i + \vartheta c_i = 2 \, \zeta c_i - \nu_i \quad ; \; \forall i \in N \quad (54)$$

Constraints (50)-(51) ensure that $\theta c_i = 1$ if the analyzed health data storage traffic traverses node $i$ to send the analyzed data from PSs at node $s$ to cloud storage $d$, otherwise it is zero. However, this does not include the last node (i.e. cloud storage) that performs the storage task. Hence, constraints (52) and (53) are to ensure $\vartheta c_j = 1$ if the traffic traverses node $j$ (including the last node) while constraint (54) is used to determine the activation of all nodes to relay and store the analyzed health data storage traffic by ensuring that the $\zeta c_i = 1$ if at least any of $\theta c_i$ and $\vartheta c_i$ are equal to 1 ($\theta c_i$OR $\vartheta c_i$), otherwise $\zeta c_i$ is zero. We achieve this by introducing a binary variable $\nu_i$ which is only equal to 1 if $\theta c_i$ and $\vartheta c_i$ are exclusively equal to 1 ($\theta c_i$ XOR $\vartheta c_i$) otherwise, it is zero.

*11)* Number of physical resource blocks at each BS to relay raw health data traffic from clinics to the fog nodes:

$$Pa_{ij} = \sum_{s\in CL} \sum_{d\in FN:s\neq d} \frac{P_{ij}^{sd}}{\delta a} \quad ; \; \forall i \in CL, \forall j \in BS: i \neq j \quad (55)$$

$$\sum_{j\in BS} Pa_{ij} = Pt_i \quad ; \; \forall i \in CL \quad (56)$$

$$\beta a_j = \sum_{i \in CL} Pa_{ij}\ Ra \quad ; \ \forall j \in BS \tag{57}$$

$$\beta a_j \le R \quad ; \ \forall j \in BS \tag{58}$$

Constraint (55) is used to ensure that each patient in the clinic is served by single BS to send the raw health data based on the traffic traversing the BS and the size of raw health data traffic of each patient. Constraint (56) is used to ensure that all patients are served by BSs. Constraint (57) calculates the total number of PRBs used at each BS. Constraint (58) is used to ensure that the number of PRBs used in each BS $j$ do not exceed the maximum number of PRBs dedicated for healthcare applications to perform the processing task.

*12)* Number of physical resource blocks at each BS to relay the analyzed health data traffic from PSs to clinics.

$$Pb_{ij} = \sum_{s \in FN} \sum_{d \in CL: s \neq d} \frac{F_{ij}^{sd}}{\delta b} \quad ; \ \forall i \in BS, \forall j \in CL \tag{59}$$

$$\sum_{i \in BS} Pb_{ij} = Pt_j \quad ; \ \forall j \in CL \tag{60}$$

$$\beta b_i = \sum_{j \in CL} Pb_{ij}\ Rb \quad ; \ \forall i \in BS \tag{61}$$

$$\beta b_i \le R \quad ; \ \forall i \in BS \tag{62}$$

Constraint (59) ensures that the analyzed health data of each patient transmitted to the clinics is relayed by single BS based on the traffic traversing the BS, and the size of analyzed health data feedback traffic of each patient. Constraint (60) ensures all patients of a clinic are served by BSs. Constraint (61) calculates the total number of PRBs used at each BS. Constraint (62) is used to ensure that the number of PRBs in each BS does not exceed its maximum number of PRBs, R, that are dedicated for healthcare applications to perform the feedback task.

*13)* Maximum number of patients served by a fog node.

$$\sum_{s \in CL} \omega_{sd} \le \Omega max\ \phi_d \quad ; \ \forall d \in FN \tag{63}$$

Constraint (63) ensures that the total number of patients served by a fog node does not exceed the number of users that can be served by the servers placed in the fog node.

*14)* Processing and analysis time at each PS.

$$\tau p_d = m \sum_{s \in CL} \omega_{sd} + \acute{c}\ \phi_d \quad ; \forall d \in FN \tag{64}$$

Constraint (64) calculates the processing and analysis time at each fog node. This is based on the total number of patients served by the PS and the number of PSs used where $m$ is the gradient of the graph while $\acute{c}$ is the *y*-intercept of the graph.

*15)* Storage capacity constraint at each PS.

$$\sum_{s \in CL} \omega_{sd}\ \alpha \le \Lambda max\ \phi_d \quad ; \ \forall d \in FN \tag{65}$$

$\forall j \in BS$

Constraint (65) ensures that the analyzed data store at a fog node does not exceed the storage capacity of the servers placed at the fog node. Note that we consider the storage capacity at the central cloud is large enough to permanently store all the analyzed data.

We compare the energy consumption of the EEFC model for the FOA to the conventional approach (CA) where the location of the PSs, $FN$ is fixed at the cloud (i.e. cloud switch). In the following, we give the MILP model for the CA (i.e. Energy efficient cloud computing (EECC) model). Note that, we used the same parameters, variables and objective function as in the EEFC model. However, as the location of the PS is at the cloud, therefore, additional variables and a modified set as in Table 2 are used in the EECC model. Also, for EECC model, we replace the word fog node used in the EEFC model to cloud node.

Table 1: Additional variables used in EECC model

| Set | |
|---|---|
| $FN$ | Set of candidate locations to deploy PS ($FN \in CLS$) |
| Variables | |
| $ECASP$ | Energy consumption of cloud data centre aggregation switches required to relay raw health data traffic |
| $ECASF$ | Energy consumption of cloud data centre aggregation switches required to relay analyzed health data feedback traffic |
| $EARP$ | Energy consumption of aggregation routers required to relay raw health data traffic |
| $EARF$ | Energy consumption of aggregation routers required to relay analyzed health data feedback traffic |
| $ECRP$ | Energy consumption of core routers required to relay raw health data traffic |
| $ECRF$ | Energy consumption of core routers required to relay analyzed health data feedback traffic |
| $ECLRP$ | Energy consumption of cloud routers required to relay raw health data traffic |
| $ECLRF$ | Energy consumption of cloud routers required to relay analyzed health data feedback traffic |
| $ECLSP$ | Energy consumption of cloud switches required to relay raw health data traffic |
| $ECLSF$ | Energy consumption of cloud switches required to relay analyzed health data feedback traffic |
| $ECSN$ | Energy consumption of cloud server node |

The energy consumption of access network, $EAN$, is the same as in Equation (3). The energy consumption of metro network, $EMN$, in Equation (15) is redefined as below:

$$EMN = (ECASP + ECASF + ECASS + EARP + EARF + EARS)\ \eta \tag{66}$$

where $ECASS$ and $EARS$ are the same as in Equation (16) and Equation (17), respectively, while others are given as:

$$ECASP = \sum_{i \in CAS} \left( ICAS\ x\ \zeta a_i + P_i \frac{PCAS - ICAS}{CCAS} \right) \tau a \tag{67}$$

$$ECASF = \sum_{i \in CAS} \left( ICAS\ x\ \zeta b_i + F_i \frac{PCAS - ICAS}{CCAS} \right) \tau b \tag{68}$$

$$EARP = \sum_{i \in AR} \left( IAR\ x\ \zeta a_i + P_i \frac{PAR - IAR}{CAR} \right) \tau a \tag{69}$$

$$EARF = \sum_{i \in AR} \left( IAR\ x\ \zeta b_i + F_i \frac{PAR - IAR}{CAR} \right) \tau b \tag{70}$$

Equations (67) and (68) depict the energy consumed by the cloud data centre aggregation switches to relay raw health data traffic, $ECASP$, and analyzed health data feedback traffic, $ECASF$, respectively. Meanwhile, Equations (69) and (70) depict the energy consumed by the aggregation routers to relay raw health data traffic, $EARP$, and analyzed health data feedback traffic, $EARF$, respectively.

The energy consumption of core network, $ECN$, in Equation (18) is redefined as below:

$$ECN = (ECRP + ECRF + ECRS)\,\eta \tag{71}$$

where the $ECRS$ is the same as in Equation (18) while other variables are given as:

$$ECRP = \sum_{i \in CR} \left( ICR\ x\ \zeta a_i + P_i\ \frac{PCR - ICR}{CCR} \right) \tau a \tag{72}$$

$$ECRF = \sum_{i \in CR} \left( ICR\ x\ \zeta b_i + F_i\ \frac{PCR - ICR}{CCR} \right) \tau b \tag{73}$$

Equations (72) and (73) depict the energy consumed by the core routers to relay the raw health data traffic, $ECRP$, and analyzed health data feedback traffic, $ECRF$, respectively.

The energy consumption of cloud in Equation (20) is redefined as below:

$$ECL = (ECLRP + ECLRF + ECLRS + ECLSP + ECLSF + ECLSS + ECSS + ECSTS)\,c \tag{74}$$

where $ECLRS$, $ECLSS$, $ECSS$ and $ECSTS$ are the same as in Equation (21)-(24), respectively, while others are given as:

$$ECLRP = \sum_{i \in CLR} \left( ICLR\ x\ \zeta a_i + P_i\ \frac{PCLR - ICLR}{CCLR} \right) \tau a \tag{75}$$

$$ECLRF = \sum_{i \in CLR} \left( ICLR\ x\ \zeta b_i + F_i\ \frac{PCLR - ICLR}{CCLR} \right) \tau b \tag{76}$$

$$ECLSP = 2 \sum_{i \in CLS} \left( ICLS\ x\ \zeta a_i + P_i\ \frac{PCLS - ICLS}{CCLS} \right) \tau a \tag{77}$$

$$ECLSF = 2 \sum_{i \in CLS} \left( ICLS\ x\ \zeta b_i + F_i\ \frac{PCLS - ICLS}{CCLS} \right) \tau b \tag{78}$$

Equations (75) and (76) depict the energy consumed by the cloud routers to relay the raw health data traffic, $ECLRP$, and analyzed health data feedback traffic, $ECLRF$, respectively. Meanwhile, Equations (77) and (78) depict the energy consumed by the cloud switches to relay the raw health data traffic, $ECLSP$ and analyzed health data feedback traffic, $ECLSF$, respectively. Note that, the energy consumption of cloud switches to transmit the traffic is multiplied by '2' for equipment redundancy purposes [32].

The energy consumption of a fog node in Equation (25) is redefined as below:

$$ECSN = EPS\ c + ETES\ \eta \tag{79}$$

where $EPS$ and $ETES$ are the same as in Equation (26) and Equations (27)-(30), respectively.

## IV. Parameter Selections

We consider two types of health monitoring application that differ in data rate. First is a low data rate ECG monitoring application. It has been reported that cardiovascular disease (CVD) has emerged as the top cause for mortality worldwide and is expected to reach 23.3 million by 2030 [4], [39]. Therefore, patients with postoperative atrial fibrillation (AF), a common cardiac case following cardiac surgery are considered in the ECG monitoring application [40]. Each patient will send a 30-second ECG signal as recommended in [40] which requires high processing capabilities for processing and analysis. The second application is a high data rate, fall video monitoring application to monitor elderly patients who suffer from heart disease. It has been reported that falls account for 10% – 25% of the ambulance call-outs for elderly people. The fall monitoring application ensures that elderly patients living by themselves get help when they experience a fall. In the event of a patient fall, the IoT device installed at the elderly home will first detect the fall using limited video processing capabilities and send a 15-second video recording as proposed in [41] to the fog servers with higher processing capabilities to reconfirm the occurrence of the fall before triggering a doctor call. Advanced processing at the fog can avoid false alarms which cost the national health service (NHS) in the UK £115 per ambulance call-out [42].

These following subsections elaborate on the methodology of determining the model input parameters considered in this work.

*1) Network layout*

In this study, 37 clinics located at West Leeds, UK, according to 2014/2015 data [25] are selected to monitor patients of the two applications. The patients of a clinic are considered to be located at the clinic due to the uncertainty in their precise locations. Potential BSs to serve patients are selected by looking into the distance between the clinics and the BSs. Note that the locations of clinics and BSs (i.e. latitude and longitude) refer to the actual locations found in West Leeds, which had been obtained from Google Maps based on the names of clinics listed by [43] in 2014/2015 and OFCOM UK Mobile Site finder published in May 2012 [44], respectively. In this work, LTE-M is considered to serve the health application with a coverage radius of less than 11km [45]. Hence, patients could be served by a BS within 11km from their registered clinics. As for this work, 315 BSs are located less than 11km from any clinic. The 26 nearest BSs to the clinics were selected to serve patients to reduce the model complexity. An OLT is selected to be placed within 20 km of the 26 BSs (the maximum distance from ONU to OLT [38], [46]) based on the location of local exchange provided by BT Wholesale network [47]. Figure 4 portrays an overview of the GPON network considered in this study.

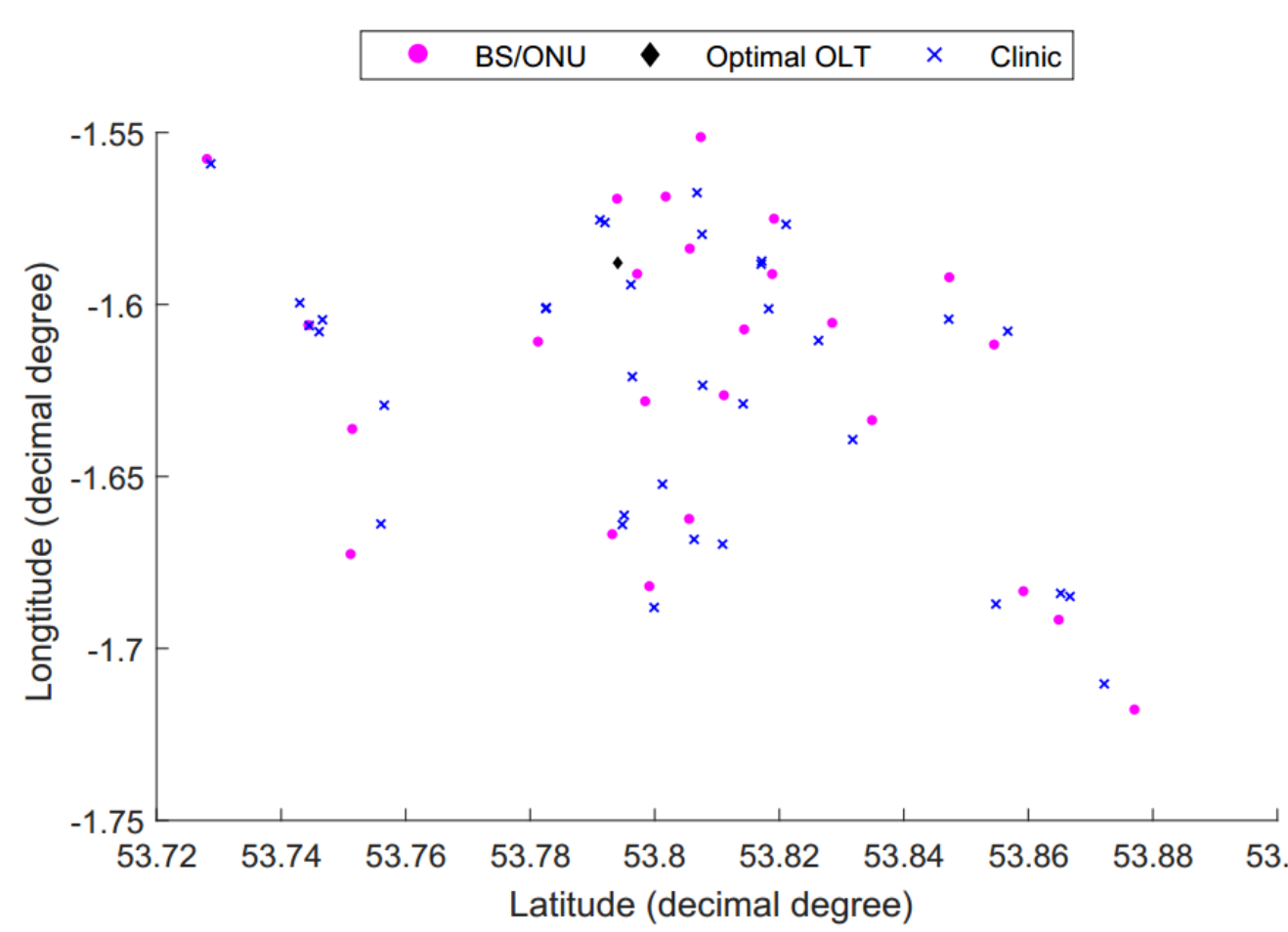

Figure 5: Locations of clinics, BSs/ONUs and OLT of the GPON network

*2) Number of monitored patients*

*a) ECG monitoring application*

According to the British Heart Foundation, the total UK population aged 18 years and above suffering from Coronary

Artery Bypass Graph (CABG) and Percutaneous Coronary Interventions (PCIs) who had heart surgeries performed in the National Health System (NHS) and selected private hospitals in 2014 were 17,513 and 96,143, respectively [48]. Given that the UK population was 63,818,387 in 2014 [49], these patients represent 0.176% of the UK population. This percentage is used to estimate the number of monitored patients in West Leeds, UK clinics based on the total number of patients registered in each clinic [43]. Table 3 presents the deduced total number of patients registered at each clinic who have been expected to experience postoperative AF.

*b) Fall monitoring application*

The total number of patients registered West Leeds clinics of all ages who suffered heart disease is obtained from Public Health England records [50]. As reported in [42], the percentage of elderly people aged 65 years and above is 17.7% of the total population and one-third of them experienced falls each year. Accordingly, 5.9% of the heart disease patients of each clinic are monitored by the fall monitoring application. Table 3 presents the deduced total number of elderly patients registered at each clinic who are expected to experience a fall.

Table 2: Number of monitored patients in clinics for ECG and fall monitoring applications

| Clinic | Number of Patients | | Clinic | Number of Patients | |
|---|---|---|---|---|---|
| | ECG | Fall | | ECG | Fall |
| Craven Road Medical Practice | 20 | 3 | Leeds Student Practice | 68 | 0 |
| Hyde Park Surgery | 18 | 1 | Burton Croft Surgery | 20 | 4 |
| Laurel Bank Surgery | 13 | 1 | Kirkstall Lane Medical Centre | 15 | 1 |
| Burley Park Medical Centre | 23 | 4 | Thornton Medical Centre | 16 | 5 |
| Gildersome Health Centre | 6 | 2 | The Dekeyser Group Practice | 30 | 8 |
| Leigh View Medical Practice | 29 | 6 | West Lodge Surgery | 32 | 13 |
| Hillfoot Surgery | 13 | 2 | Dr. KW McGechaen & Partner | 8 | 2 |
| Pudsey Health Centre | 13 | 4 | Robin Lane Medical Centre | 24 | 6 |
| Dr. S M Chen & Partner | 8 | 2 | Beech Tree Medical Centre | 4 | 1 |
| Hawthorn Surgery | 10 | 3 | Priory View Medical Centre | 16 | 6 |
| High Field Surgery | 14 | 3 | Abbey Grange Medical Centre | 16 | 4 |
| Vesper Road Surgery | 11 | 2 | Fieldhead Surgery | 10 | 1 |
| Manor Park Surgery | 27 | 7 | The Highfield Medical Centre | 9 | 2 |
| Dr. G Leeds & Partners | 25 | 4 | Dr. F Gupta's Practice | 6 | 1 |
| Guiseley and Yeadon Medical Practice | 21 | 6 | Park Road & Menston | 19 | 6 |
| Yeadon Tarn Medical Practice | 12 | 4 | Rawdon Surgery | 14 | 4 |
| Dr. KJ Manock & Partners | 44 | 11 | Whitehall Surgery | 16 | 2 |
| Dr. JA Browne's Practice | 28 | 6 | Dr. N Saddiq's Practice | 5 | 1 |
| Dr. JJ McPeakes Practice | 6 | 2 | | | |

*3) Link capacity*

The M2M traffic was 2% of the global IP traffic in 2016 and is expected to be 5% in 2021 [51]. Cisco also reported that the connected health applications will represent 6% of M2M traffic in 2020 [52]. Accordingly, healthcare applications are estimated to account for 0.3% of the global IP traffic. This percentage is used to estimate the network link capacities available for healthcare applications.

*4) Time for processing and analysis*

*a) ECG monitoring application:*

For the ECG monitoring application, a 30-second ECG signal, illustrated in Figure 6, is required to be sent to monitor postoperative AF of cardiac surgical patients. This signal is retrieved from the MIT_BIT Arrhythmia database [53], [54]. Note that, the 30-second ECG signal offers accurate results for the analysis, as recommended in [40]. Such 30-second of un-processed ECG signals have a volume of 252.8 kbits. The ECG signals is processed using Pan Tompkins algorithm [43] to extract heart rate and QRS duration for further analysis. The calculation of the heart rate from the 30-second ECG signal is based on the number of R waves within the 30 seconds and this number is multiplied by 2 to obtain the heart rate in beats per minute [55]. The QRS duration is obtained based on the time between Q and S waves found in the ECG signal.

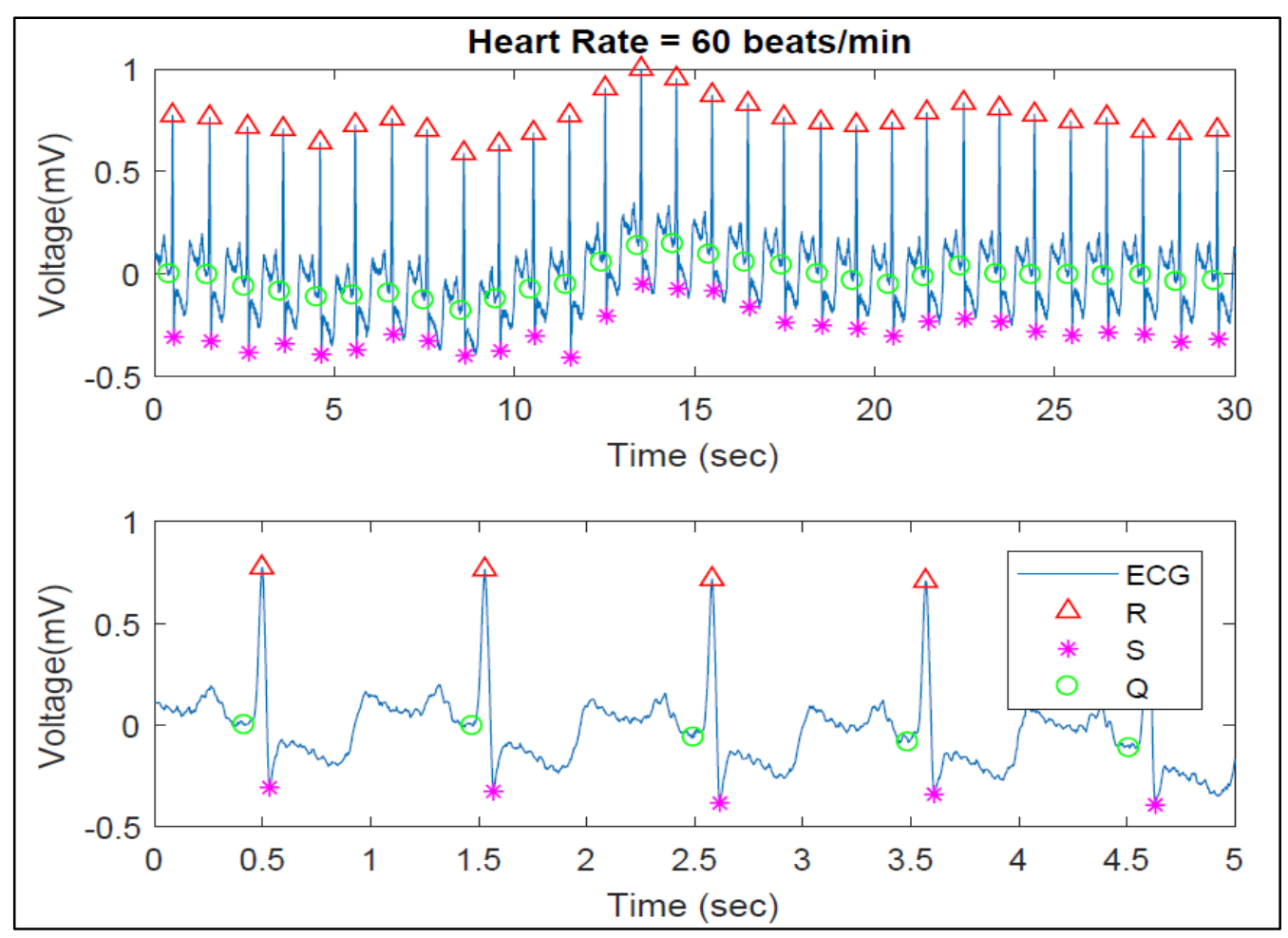

Figure 6: The 30-second ECG waveform

The PS selected in both fog and central cloud to process ECG signal is Intel Core i5-4460 with 3.2 GHz CPU and 500 GByte hard drive [56]. An experiment was conducted using MATLAB with a parallel processing function to determine the correlation between time and number of patients for processing and analysis of raw ECG data using Pan-Tompkins algorithm. This was carried out by performing the processing task on the 30-second ECG signals generated by 10k to 50k patients in 10k steps. At each 10k step, the processing operation was repeated 5 times to calculate the average time for the processing duration. Note that, the 30-second ECG signals are made up of 1 ECG record repeated for all patients. Also, note that the time to perform the processing using MATLAB consists of both the time to submit the data for parallel processing and the time to run the algorithm.

The results were then fitted with a linear line (dotted blue line), as illustrated in Figure 7. For instance, a 10-second duration for processing could process 2657 patients. We also obtained the correlation between the time and number of patients for the processing and analysis of raw ECG signal considering 41 ECG records retrieved from the MIT_BIT Arrhythmia database [53], [54] with a duration of 30-seconds each. The patient's ECG signals are uniformly selected from these 41 records. The results are as shown as a red line in Figure 7. The two experiments with a single ECG signal and multiple ECG signals have resulted in similar linear relationships. The MILP model results in the next section are obtained considering the single ECG signal. Therefore, for ECG monitoring applications;

$$\tau p = 0.002\ Pat + 4.6872 \quad (80)$$

where $Pat$ is the number of patients served in each PS.

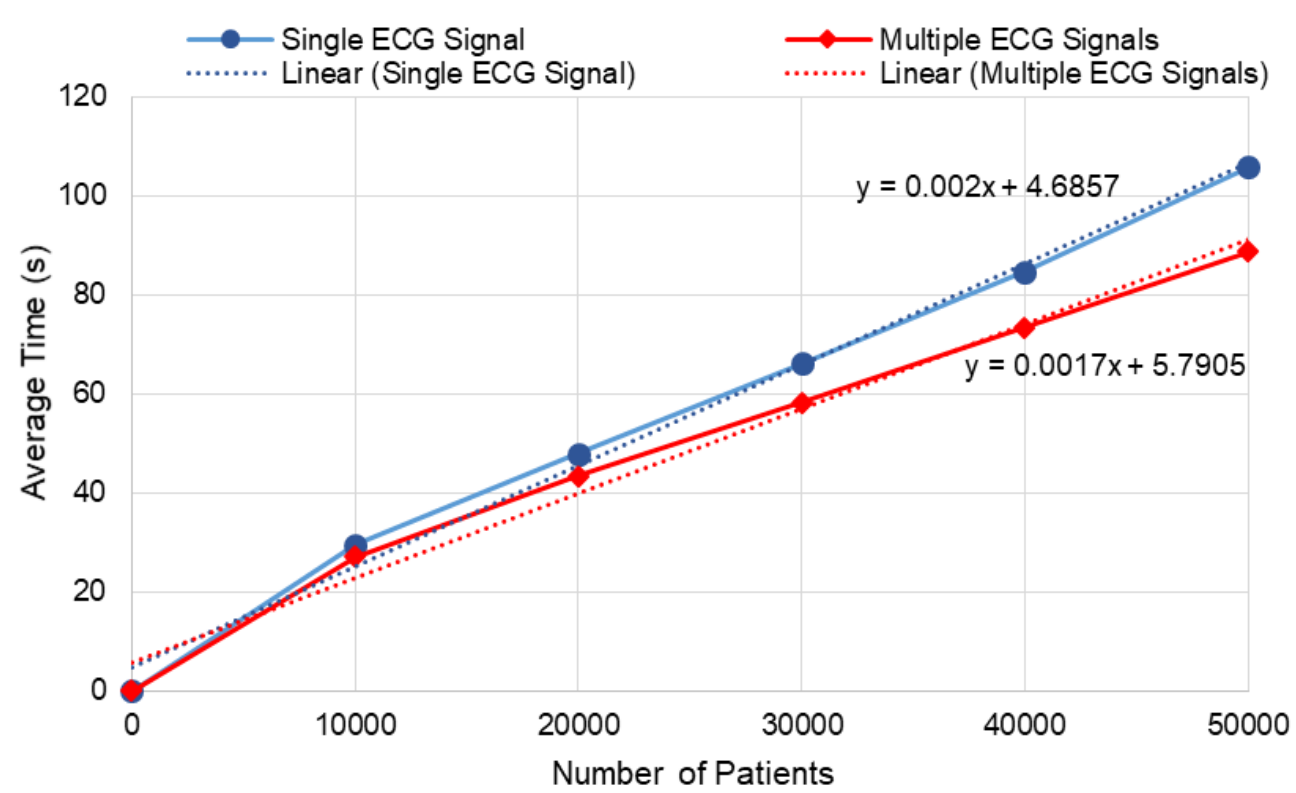

Figure 7: Number of patient versus time, based on MATLAB simulations

*b) Fall monitoring application*

For the fall monitoring application, the 15-second video recording is captured using a Kinect's IR sensor with a 640 × 480 resolution at 30 frames per second, as proposed in [41]. Therefore, the video transmitted by a patient is 3.36 Mbits. Note that, the PS selected in both fog and cloud for fall monitoring application is 2.4 GHz Intel Core-Duo. The time to process and analyze the video with a 2.4-GHz processor is around 0.3 ms – 0.4 ms per frame [41]. In this work, 0.4 ms was used as the per frame processing time. Therefore, the duration to process and analyze one video recording per patient is 0.18 s, as calculated below:

$$\tau ps = 15\ s\ \ 30\ frames/s\ 0.4\ ms/frame \quad (81)$$

In this work, videos were assumed to be processed in series. Therefore, the worst-case scenario was considered to be one in which all the videos are processed and analyzed before the feedback was sent. Hence, for fall monitoring application,

$$\tau p = \tau ps\ Pat \quad (82)$$

where $\tau ps$ is the duration to process a video recording per patient.

*5) Patient Data rate*

The American Heart Association (AHA) has recommended that the golden time to save a heart patient's life by sending an alarm message to a cardiologist upon detection of a sudden fall or rise in cardiac vital signs is between 4 and 6 minutes [57]. As such, 4 minutes, $\tau t$, was selected for this work as the maximum duration to perform all tasks which is composed of: i) the time to record the health data, $\tau m$, given as 30 seconds for ECG monitoring application while 15 seconds for fall monitoring application as explained above ii) the time to transmit raw health data to the PS, $\tau max$, iii) the time for processing and analysis of raw heal data, $\tau p$, estimated as explained above and iv) the time to transmit the analyzed health data to clinic for feedback , $\tau b$. Therefore, latency is not considered in this work as the time to perform the main tasks explained above to save the heart patients is limited to 4 minutes. Note that, this time frame is used to calculate the minimum data rate for each patient.

The time required to transmit the analyzed data to the clinics for feedback, $\tau b$, is determined based on the GPON link bandwidth available for the processing node ( fog node or cloud) to send feedback data to clinics ($Cb_{min}$), and the amount of feedback data each PS needs to send. The link bandwidth available for the processing node is estimated as 0.3% of the uplink and downlink between the ONU and the OLT for FOA and CA, respectively, as explained above.

Note that we choose to transmit the feedback data using the maximum data rate available for healthcare applications in the GPON links to decrease the feedback time which, in turn, gives more time to transmit the raw health signal which allows a lower data rate. This will result in activating fewer BSs. Note that, activating fewer BSs for a longer time is more efficient than activating a large number of BSs for a shorter time as the idle power consumption of a BS is 63% of its total power. The feedback data rate transmitted by a fog node is calculated considering the maximum number of patients that can be served at a candidate node, $Pat_{max}$, given as:

$$Pat_{max} = Pat\ N \quad (83)$$

where $N$ is the number of PSs that can be hosted at a candidate fog node.

We divided the minimum link/device capacity ($Cb_{min}$) equally among patients, hence the data rate available for each patient, $\delta f$, is given as:

$$\delta f = Cb_{min}/Pat_{max} \quad (84)$$

In this study, an LTE-M base station with the QPSK modulation scheme is considered with a minimum of 336 bps per physical resource block (PRB). Therefore, the number of PRBs for each patient to send the feedback data is given as:

$$Rb = \lfloor \delta f/336\ bits \rfloor \quad (85)$$

where $Rb$ is the minimum integer value to ensure that the link capacity that was provisioned for healthcare in the network was not exceeded. Therefore, the data rate to send the feedback data is given as:

$$\delta b = Rb\ 336\ bits \quad (86)$$

while the transmission time is calculated as follows:

$$\tau b = \alpha/\delta b \quad (87)$$

where $\alpha$ is the size of the analyzed data (feedback data). For ECG monitoring application, the size of analyzed health data ($\alpha$) to be sent to the clinics for feedback purposes and to be permanently stored in the cloud storage is obtained from the conducted experiment, which is 256 bits. Meanwhile, for fall monitoring applications, the size of analyzed data is based on the maximum size allowed for a notification payload according to Apple Push Notification Services, which is 256 bytes (i.e. 2.048 kbits) [58].

Therefore, the transmission time to send the raw health data to the PSs is given as:

$$\tau max = \tau t - \tau m - \tau b - \tau p \quad (88)$$

The minimum data rate to transmit the raw health data to the PSs is calculated based on $\tau max$ as:

$$\delta min = D/\tau max \quad (89)$$

where $D$ refers to the size of the raw health data. However, as the data traverses the LTE base station and the minimum allocation of resources to each user was one $PRB$, the number of PRBs that could be assigned to each patient to transmit his/her raw health data is:

$$Ra = \lceil \delta min/336\ bps \rceil \quad (90)$$

where $Ra$ is the maximum integer value to ensure that the given data rate is equal to or higher than the minimum required data rate so that the system could work within 4-minute. Hence, the data rate to send raw health data to the PS is:

$$\delta a = Ra\ 336\ bps \quad (91)$$

while the transmission time to send raw health data is calculated as below:

$$\tau a = D/\delta a \quad (92)$$

The data rate to send analyzed health data from PSs to the cloud storage for permanent storage is given as:

$$\delta c = Cc_{min}/Pat_{max} \quad (93)$$

where$Cc_{min}$ is the lowest shared uplink or node capacity available for a health M2M application from the PS to the cloud storage (i.e. uplink capacity between ONU and OLT and content server capacity for FOA and CA, respectively). Hence, the time required to transmit the analyzed health data to cloud storage is:

$$\tau c = \alpha/\delta c \quad (94)$$

*9) Equipment power consumption*

As explained in Section III, the power consumption of all networking equipment and PS consist of an idle part and a linear proportional part. The idle power of BS, PS, and content server are obtained from datasheets and references in [7], [59] and [60], respectively while the idle power for the other networking devices was considered to be 90% of the power consumption at maximum utilization [7], [61] and [62]. The maximum power consumption of the networking equipment and the PS and their maximum capacity is given by the manufacturers. As for ONU, the maximum capacity, $CONU$, is considered as the summation of the maximum uplink capacity, i.e. 1.25 Gbps [63] and maximum downlink capacity, i.e. 2.5 Gbps [63], to obtain $E_b$. Note that, the networking devices are shared by multiple applications while the considered PSs and Ethernet switch are dedicated for the healthcare application. As discussed for the link capacity, in this work we consider 0.3% of the idle power of the shared devices is contributed by our healthcare applications while 0.42% for LTE-M BS. Note that, the LTE-M shares capacity, antenna, radio, and hardware with the legacy LTE networks (20MHz) [45]. Due to this, the calculated idle power of the BS (0.42%) contributed for healthcare applications is based on 7% allocation of LTE-M network from the legacy LTE network (i.e. 1.4MHz/20MHz) and 6% [52] allocation of healthcare application from the total M2M application supported by LTE-M network. Note that, the 6% allocation refers to the estimated total number of RBs that is dedicated for healthcare applications which gives 360 PRBs per second as there are numerous types of M2M applications served by LTE-M. However, the maximum idle power is considered for the unshared devices.

Due to cooling, lighting and other overheads in the network, the total power consumed in a site is higher than the power consumed by the communications and computing equipment. The ratio of the total power consumed to the power consumed by the communications and computing equipment is defined as the power usage effectiveness (PUE). PUE is used to describe the energy efficiency of each site (core node site or building, cloud site or building or fog site). A PUE of 1.5 is considered for IP over WDM, metro, and access networks [64], [65]. A PUE of 2.5 is considered for small distributed clouds in this work [32]. In addition, a PUE of 2.5 is set for fog. Table 4 depicts the input parameters of the models for the network architecture.

Table 3: Input parameters for networking and computing devices

| Parameter | Value |
|---|---|
| Maximum power consumption of core router (CRS-3), $PCR$ | 12300 W [61] |
| Core router capacity (CRS-3), $CCR$ | 4480 Gbps [61] |
| Maximum power consumption of cloud switch (Catalyst 6509), $PCLS$ | 2020 W [61] |
| Cloud switch capacity (Catalyst 6509), $CCLS$ | 320 Gbps [61] |
| Maximum power consumption of cloud router (7609), $PCLR$ | 4550 W [61] |
| Cloud router capacity (7609), $CCLR$ | 560 Gbps [61] |
| Maximum power consumption of content server, $PCS$ | 380.8 W [60] |
| Idle power consumption of content server, $ICS$ | 324.82 W [60] |
| Content server capacity, $CCS$ | 1.8 Gbps [60] |
| Maximum power consumption of cloud storage, $PCST$ | 4900 W [64] |
| Cloud storage capacity $CCST$ | 75.6 TB [64] |
| Maximum power consumption of aggregation router (7609), $PAR$ | 4550 W [7], [61] |
| Aggregation router capacity (7609), $CAR$ | 560 Gbps [7], [61] |
| Maximum power consumption of cloud data centre aggregation switch, (Catalyst 6509), $PCAS$ | 1766 W [61] |
| Cloud data entre aggregation switch capacity (Catalyst 6509), $CCAS$ | 256 Gbps [61] |
| Maximum power consumption of OLT, $POLT$ | 20 W [46] |
| OLT capacity, $COLT$ | 128 Gbps [46] |
| Maximum power consumption of ONU, $PONU$ | 8 W [63] |
| ONU capacity, $CONU$ | 3.75 Gbps |
| Maximum power consumption of LTE Base Station, $PBS$ | 528 W [66] |
| Idle power consumption of LTE Base Station, $IBS$ | 333 W [66] |
| Maximum power consumption of Ethernet switch, $PES$ | 3.52 W [67] |
| Idle power consumption of Ethernet switch, $IES$ | 16 Gbps [67] |
| Ethernet switch capacity, $CES$ | 0.57 W [67] |
| Maximum power consumption of PS, $PPS$ | 180 W [59] |
| Idle power consumption of PS, $IPS$ | 78 W [59] |
| IP over WDM, access and metro network PUE, $\eta$ | 1.5 [64], [65] |
| Cloud and fog PUE, c | 2.5 [32] |

## V. Performance evaluation for the ECG monitoring application

This section presents the results and analysis of the EEFC model for the proposed fog optimized approach (FOA) and the EECC model for the conventional approach (CA) considering the ECG monitoring application. AMPL software with CPLEX 12.8 solver running on high-performance computing (HPC) cluster with a 12 core CPU and 64 GB RAM was used as a platform for solving the MILP models. The performance of the EECC model where the ECG signals are processed in the cloud is used as a benchmark to evaluate the performance of the EEFC model in terms of energy consumption of networking equipment and processing. Note that the evaluation of the two models is performed using the GPON architecture, as shown in Figure 1. In addition, a heuristic, Energy Optimized Fog Computing (EOFC), is also introduced for real-time implementation of the proposed approach based on insights from the MILP results. The performance of the EOFC heuristic running on a normal personal computer (PC) with 3.2 GHz CPU and 16GB RAM is also evaluated. Also, the impact of the idle power of the servers and network devices and traffic volume on the energy savings achieved by the proposed approach (i.e. FOA) is evaluated.

*1)* The MILP model with GPON network:

Based on the outcomes of the MILP model, the energy consumed by both networking equipment and processing in GPON network had been determined via two approaches: CA and FOA. Table 5 shows the calculated input parameters for the FOA and CA for ECG monitoring applications. Note that, we consider a single PS to serve all patients ($Pat = 669$ patients). Also, we consider a scenario where we only allow one PS at each candidate node ($N = 1$) as the limited space at the fog node can be shared by multiple applications, i.e. $\phi_d$ will be a parameter (i.e. $\phi_d = 1$).

Table 4: Input Parameters for FOA and CA for ECG monitoring applications when $N = 1$ and a single PS can serve all patients.

| Parameter | FOA | CA |
|---|---|---|
| Size of ECG data (kbits) | 252.8 | 252.8 |
| Size of analyzed ECG data (bits) | 256 | 256 |
| Transmission time to transmit ECG data to PS, $\tau a$ (s) | 188.1 | 188.1 |
| Data rate to transmit ECG data to PS, $\delta a$ (bps) | 1344 | 1344 |
| Transmission time to transmit analyzed ECG data to clinic, $\tau b$ (s) | 0.76 | 0.38 |
| Data rate to transmit analyzed ECG data to clinic, $\delta b$ (bps) | 336 | 672 |
| Transmission time to transmit analyzed ECG data to cloud storage, $\tau c$ (s) | 0.73 | 0.032 |
| Data rate to transmit analyzed ECG data to cloud storage, $\delta c$ (bps) | 350 | 8070 |

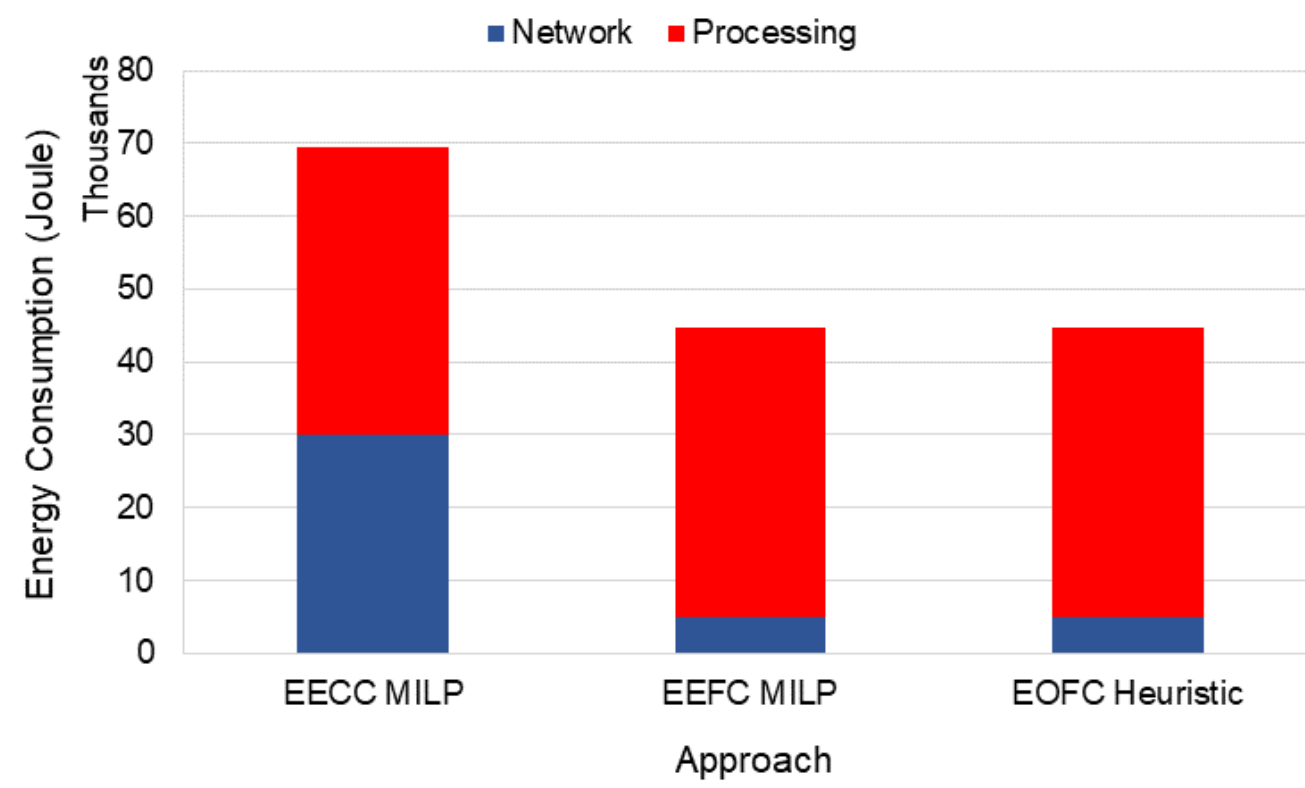

Figure 8: Energy consumption of networking equipment and processing in GPON architecture

Figure 9 shows the energy consumption of networking equipment and processing for the EECC model and EEFC model. The energy saving of networking equipment achieved by the EEFC model compared to the EECC model is 83.1%, as illustrated in Figure 9. The MILP results show that there is only one PS deployed at the OLT as it is the nearest shared point to the patients (the OLT is connected to all BSs in the network). Processing the raw health data at the fog server limited the network journey of this data to the GPON, i.e. only the feedback data and permanent storage data (i.e. processed data) is sent to the cloud, resulting in reducing the metro and core network energy consumption by reducing the data traversing the network and reducing the utilization time of the network equipment, i.e. reducing the idle power consumption. Note that the larger the size of the data and the longer the transmission duration, the higher the energy consumption. Comparing that to EECC model, higher energy is consumed by the networking equipment in the metro and core layers in the EECC model as the un-processed data is sent to the central cloud to be processed. Note that permanent storage is also performed in the EECC model after the data is processed.

Figure 9 illustrates that the energy consumption for processing in the EEFC model is slightly higher than the EECC model by 0.5%. This is due to the longer time required to send the analyzed data for feedback and permanent storage from fog servers as access network link capacity is limited (336 bps and 350 bps, respectively) compared to the cloud data centre network (672 bps and 8070 bps, respectively). As shown in Table 5, under the EEFC model, the PS is on idle mode for 0.76 s and 0.73 s to send the analyzed data for feedback and permanent storage, respectively while only 0.38 s and 32 ms are required to send the data from a cloud server under the EECC model. However, the total energy saving that includes the networking equipment and processing achieved by the EEFC model compared to the EECC model is 35.7%.

*2)* The Energy optimized fog computing (EOFC) heuristic:

The Energy optimized fog computing (EOFC) heuristic is developed as a method to validate the MILP model and to deliver a real-time solution of the FOA, as the MILP solution is exponential in time with increase in the network size. The heuristic determines the BSs to serve patients to send raw health data and receive feedback data; and the access network nodes to place PSs so that the energy consumption of both networking and processing is minimized. Figure 10 shows the flow chart of the EOFC heuristic.

The heuristic begins by sorting the clinics based on the number of patients the clinic serves in ascending order. The heuristic assigns the clinic with the smallest number of patients to BSs to help packing the BSs (packing is optimum when equipment has high idle power consumption). The assignment of clinic patients to a BS is as follows: The heuristic sorts the BSs that have a connection to the clinic under consideration starting with BSs previously used by the healthcare application that has available resources. These BSs are sorted in ascending order based on the total number of clinics the BS can serve followed by the unused BSs in descending order. Ascending order of activated BSs reduces the number of utilized BS while the descending order of unused BSs ensures that options are left open until late in the allocation process. Then, the patients of the clinic under consideration are consolidated to the minimum number of BSs to reduce the number of BSs used by the healthcare application.

The heuristic then determines the number of PSs required to serve the patients and the nodes hosting them. The candidate nodes that can host the servers are the ONUs connected to the BSs selected to serve the patients and the OLTs. Considering the minimum number of nodes required to host servers to serve all the patients (which is based on the maximum number of servers a node can host), the heuristic finds the combination of candidate nodes to host PSs that result in minimum energy consumption. The aim of limiting the number of nodes to place the PSs is to reduce the utilization of the Ethernet switches to serve the PSs.

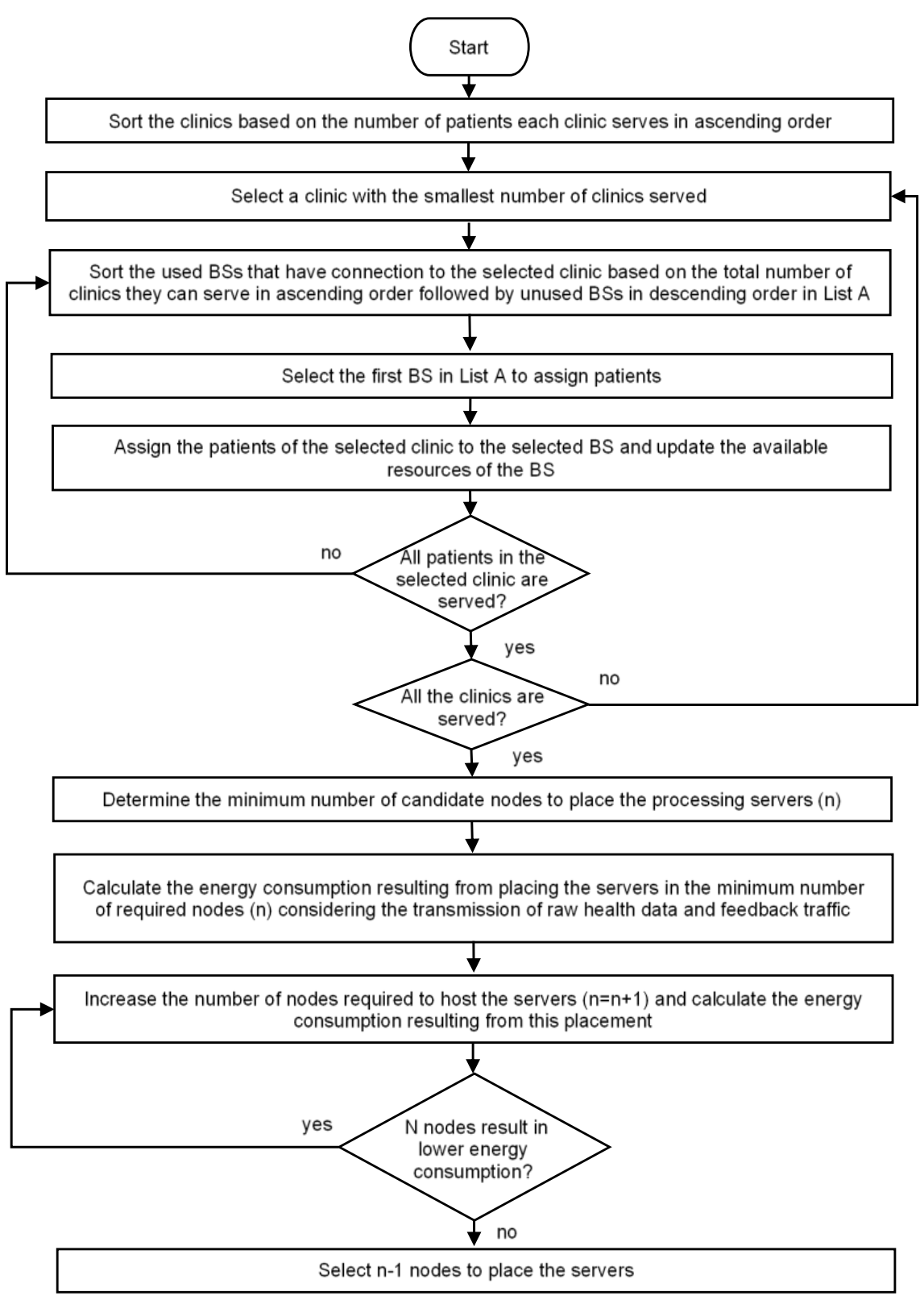

Figure 9: Flow chart of EOFC heuristic

This energy consumption that results from hosting servers at a combination of candidate nodes is calculated by routing the traffic (raw health data traffic) from BSs (starting with the BS serving the largest number of patients) to the nearest node with available processing capacity out of the combination of candidate nodes under consideration based on minimum hop routing. Also, BSs to send feedback traffic from the combination of candidate nodes to clinics are selected using the same approach used to select BSs to send raw health data. Note that BSs different from those used to send raw health data are used to send feedback traffic as the size of the analyzed health data feedback traffic is smaller than the raw health data traffic. The combination of nodes hosting servers considering the minimum number of nodes required to host servers to serve all the patients that result in minimum energy consumption is selected.

The heuristic increases the number of candidate nodes to host servers and repeats the above process. The energy consumption resulting from using this combination of nodes is calculated and compared to the energy consumption resulting from the combination of nodes hosting servers considering the minimum number of nodes required to host servers. If the latter is lower, the heuristic examines placing servers in more candidate nodes. If the former is lower, the minimum number of nodes required to host servers is selected to place servers.

The performance of the EOFC heuristic is compared to the EEFC model and the results in Figure 9 show that the EOFC heuristic has the same performance as the EEFC model in terms of the networking and processing energy consumption in comparison to the EECC model. This is because the optimal location to place the PS in both EOFC heuristic and EEFC model is at the OLT, and the same networking equipment is used to serve the patients.

*3)* Impact of idle power of devices:

We also studied the impact of the idle power of networking and processing equipment on the energy savings achieved by the EEFC approach. We reduced the idle power of all equipment (given in Table 4) by 30%, 60% and 100%. Note that because the idle power of BS, PS, and content server are obtained from [7], [59] and [60], while the other equipment is considered to be 90% of the maximum power, therefore to obtain an equivalent reduction ratio for all equipment, we considered reductions by 33%, 67% and 100% from their fixed idle power. Also, we used the same input parameters as in Table 5.

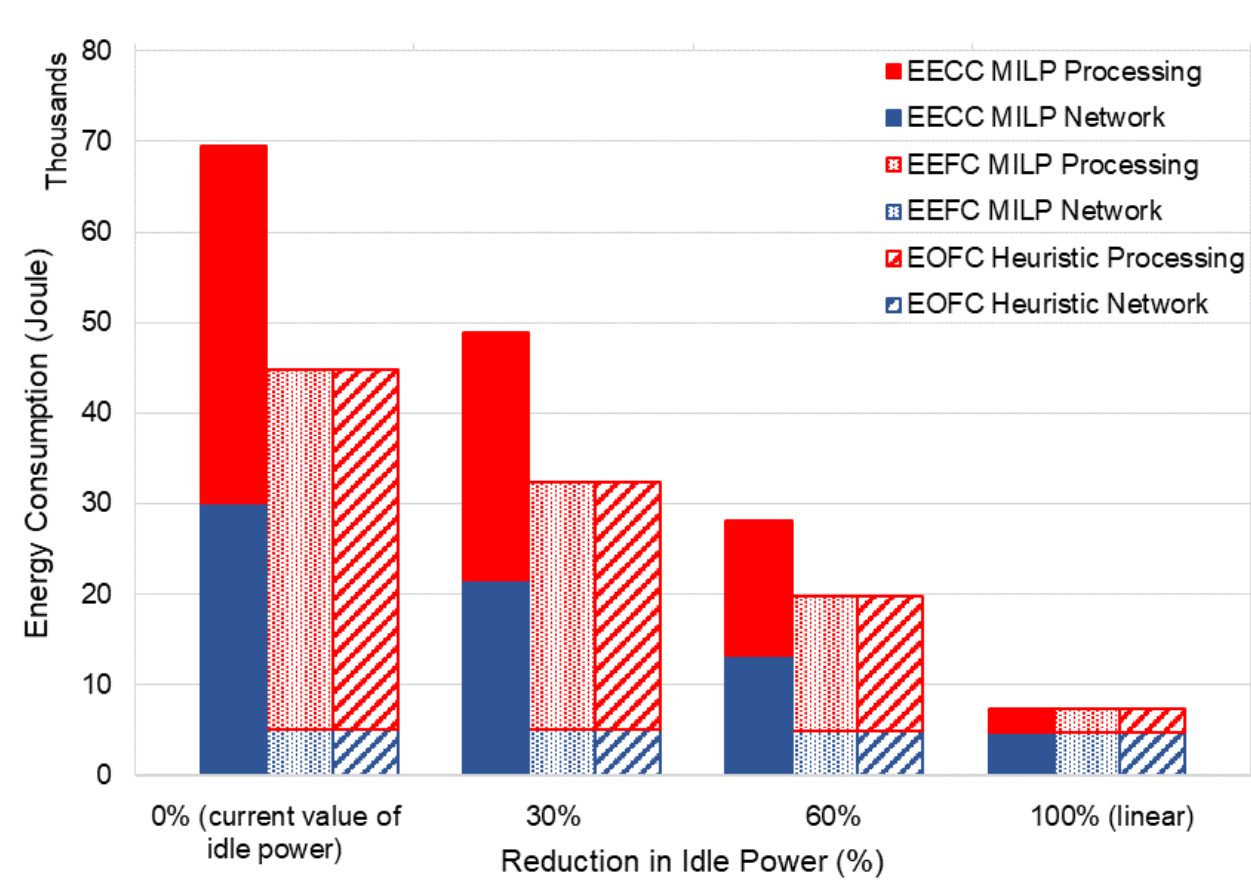

Figure 10: Energy consumption of networking equipment and processing of EECC model, EEFC model and EOFC heuristic with varying idle power consumption

Figure 11 illustrates the energy consumption of networking equipment and processing for the EECC model, EEFC model and EOFC heuristic with different idle power consumption. Figure 11 also shows that the energy consumption of networking equipment and processing in the EOFC heuristic are the same as in the EEFC model. The energy consumption of networking equipment and processing for both the EEFC model and the EECC model decrease as the idle power consumption decreases. This is because the idle power dominates the energy consumption of networking equipment and PS compared to its proportional load power as the size of data used in this work is small.

Meanwhile, the energy consumption of processing in the EEFC model and the EECC model are the same when the idle power is 0%. This is because the PS in the EEFC model and the EECC model served the same number of patients with the same processing and analysis time. Table 6 summarized the energy saving of networking equipment and the energy increase for processing in the EEFC model when compared to the EECC model for all percentages of idle power.

The results also show that the energy savings of networking equipment obtained by the EEFC model compared to the EECC model decrease with decreasing the idle power consumption. This is because, as explained in Section V-1, the EEFC savings are obtained as a result of reducing the data traversing the metro and core networks and consequently the utilization time of the metro and core network equipment, i.e reducing the idle power contribution of the healthcare application. Therefore, decreasing the idle power consumption reduces the margin of savings. Note that the decrease of idle power only affected the energy consumed due to receiving the raw ECG signal from patients and the energy consumed due to transmitting the analyzed data for feedback and permanent storage purposes as shown in Equation (26).

Table 5: Energy-Saving and Energy-Increase in FOA compared to CA, with varied percentages of idle power

| Type of energy | Percentage of Idle Power | | | |
|---|---|---|---|---|
| | 90% | 60% | 30% | 0% |
| Network Saving | 83.1% | 77.1% | 63.5% | 0.3% |
| Processing Increase | 0.53% | 0.52% | 0.47% | 0% |

*4)* The impact of increasing traffic on EEFC:

In this section, the impact of increasing the traffic on the energy consumption of networking equipment and processing in the EEFC model is evaluated by increasing the number of patients by 10% to 90% of the number of patients for each clinic in 2014/2015 in 10% increments. Note that increasing number of patients increases the traffic in the network. To maintain the processing and analysis time of each PS at 6.02s where the maximum number of patients a PS can serve ($Pat$) is equal to 669 (2014/2015 total number of patients), multiple servers will be required to serve the increasing number of patients.

We consider two scenarios related to the number of PSs that can be deployed at each candidate node (i.e. fog node). In the first scenario (Scenario 1), each candidate node can serve only one PS, hence the $\phi_d$ is a parameter (i.e. $\phi_d = 1$). The first scenario is applicable for the EEFC model only. In the second scenario (Scenario 2), each candidate node can serve more than one PS, hence $\phi_d$ is a variable. The second scenario is applicable for both the EECC model and the EEFC model. Note that, in Scenario 2, to allow each candidate node to host more than one PS, an additional networking equipment, Ethernet switch, dedicated for healthcare applications is required to connect the PSs to each candidate node

The same MILP model in Section III for CA and FOA is used to evaluate the performance of the proposed approach under GPON network. Similar input parameters to those in Table 5 are considered for the GPON network for Scenario 1 to evaluate the performance of the EEFC model in terms of energy consumption of networking equipment and processing versus increasing traffic. For Scenario 2, similar input parameters to those in Table 5 are considered, except the data rate for permanent storage ($\delta c$) and its transmission time ($\tau c$), are employed for the EECC model and EEFC model. This is because, in Scenario 2, the data rate per patient to send the analyzed data to the cloud for permanent storage for the EECC model (i.e. CA) and EEFC model (i.e. FOA) decreases with increasing number of patients (more than one PS can be served at each candidate node), which, in turn, increases its transmission time. Increasing the number of patients in the network also reduces the data rate for feedback ($\delta b$) and increases its transmission time ($\tau b$) for the EECC model (i.e. Scenario 2) to 336 bps and 0.76 ms, respectively. The values remain the same for all percentages of patients as the allocated data rate is the minimum rate in the LTE when using the QPSK modulation scheme. Table 7 shows the data rate for permanent storage, $\delta c$, and its transmission time, $\tau c$, for the EEFC model (i.e. FOA) and EECC model (i.e. CA) for Scenario 2 for increasing number of patients in the network.

Table 7: Data rate and transmission time for permanent storage with varying numbers of patient in the network for the EECC and EEFC model under Scenario 2

| Approach | CA | | FOA | |
|---|---|---|---|---|
| | δc (kbps) | τc (s) | δc (kbps) | τc (s) |
| 10% | 7.317 | 0.035 | 0.317 | 0.81 |
| 20% | 6.708 | 0.038 | 0.291 | 0.88 |
| 30% | 6.199 | 0.041 | 0.269 | 0.95 |
| 40% | 5.775 | 0.044 | 0.250 | 1.02 |
| 50% | 5.346 | 0.048 | 0.232 | 1.1 |
| 60% | 5.037 | 0.051 | 0.218 | 1.17 |
| 70% | 4.741 | 0.054 | 0.205 | 1.25 |
| 80% | 4.492 | 0.057 | 0.194 | 1.31 |
| 90% | 4.245 | 0.060 | 0.184 | 1.39 |

Due to the complexity of evaluating the MILP model for a large network for the increasing percentages of traffic, the EOFC heuristic is used to study the performance of the energy consumption of networking equipment and processing for the EEFC model compared to the EECC model.

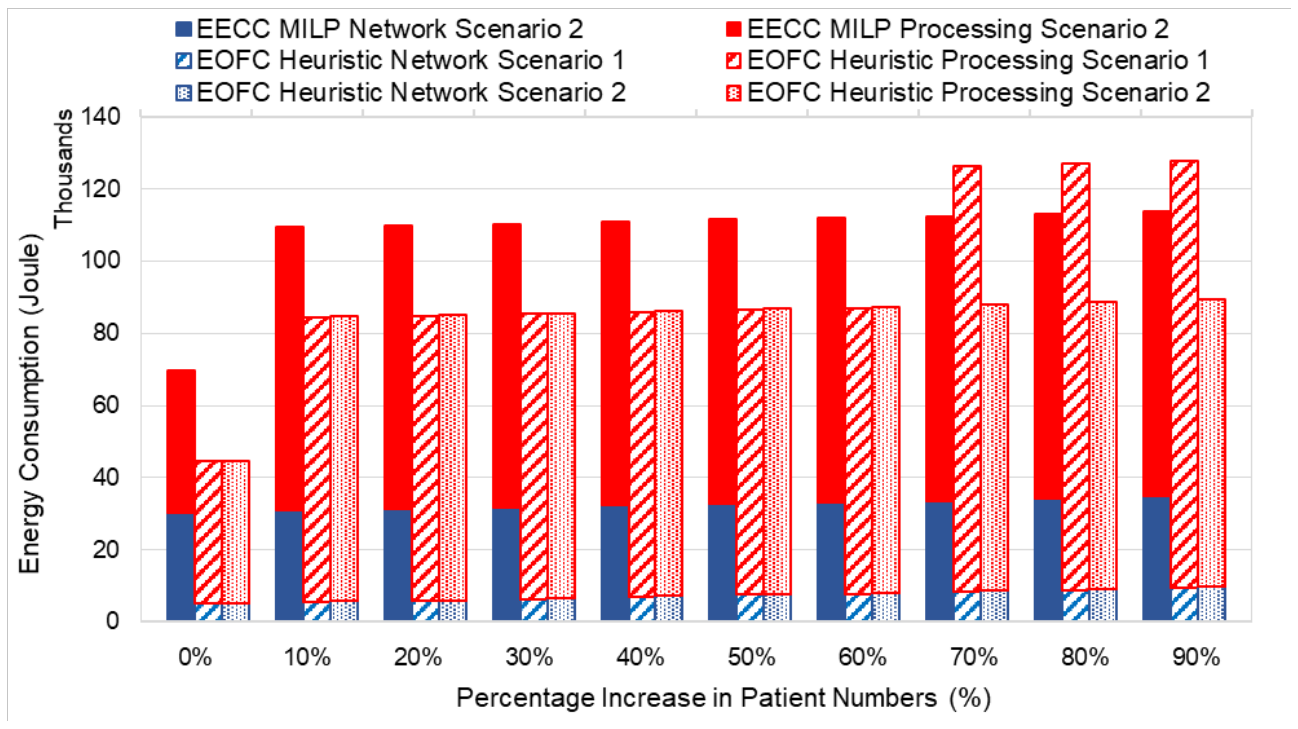

Figure 11: Energy consumption of networking equipment and processing in EOFC heuristic under Scenario 1 and EECC model and EOFC heuristic under Scenario 2 with the increasing number of patients

Table 8: Network energy savings and processing energy increases of the EOFC heuristic under Scenario 1 and Scenario 2 compared to the EECC model under Scenario 2, for varying percentages of traffic increase

| Percentage of Increasing Traffic (%) | Scenario 1 | | Scenario 2 | |
|---|---|---|---|---|
| | Network Saving (%) | Processing Increase (%) | Network Saving (%) | Processing Increase (%) |
| 0 | 83.1 | 0.53 | 83.1 | 0.53 |
| 10 | 81.7 | 0.34 | 81.1 | 0.38 |
| 20 | 81.3 | 0.34 | 80.6 | 0.42 |
| 30 | 79.8 | 0.34 | 79.1 | 0.45 |
| 40 | 78.4 | 0.34 | 77.7 | 0.48 |
| 50 | 77.0 | 0.34 | 76.3 | 0.52 |
| 60 | 76.6 | 0.33 | 75.9 | 0.55 |
| 70 | 75.3 | 33.27 | 74.5 | 0.59 |
| 80 | 74.1 | 33.25 | 73.2 | 0.62 |
| 90 | 72.8 | 33.23 | 71.9 | 0.65 |

Figure 12 shows the energy consumption of networking equipment and processing for EOFC heuristic for Scenario 1 and EECC model and EOFC heuristic for Scenario 2, when the traffic is based on 2014/2015 (i.e. 0% increase) and increased by 10% to 90% from the total number of patients for each clinic in 2014/2015 in 10% step units. Table 8 shows the network energy savings and processing energy increases of the EOFC heuristic under Scenario 1 and Scenario 2 compared to the EECC model under Scenario 2, for traffic increase between 0% and 90%.

The results in Figure 12 show that the total energy consumption of the EECC model (i.e. Scenario 2) and the EOFC heuristic (i.e. Scenario 1 and Scenario 2) increase as the traffic increases. The increase in energy is a result of increasing the number of patients which, in turn, increases both the total traffic traversing the network and the total number of utilized networking and processing equipment. The results in Figure 12 show that the total energy consumption of both networking equipment and processing in the EOFC heuristic of Scenario 1 is lower than the EECC model when the percentage increase in patients is equal to or less than 60%. Meanwhile, for the EOFC heuristic under Scenario 2, the total energy is lower than the EECC model for all percentages of increasing traffic. The low total energy consumed in the EOFC heuristic of both scenarios is mainly due to the low energy consumed by the networking equipment as a result of the small number of utilized networking equipment and its utilization time in the EOFC heuristics compared to the EECC model as explained previously

The results also show that the total energy consumed by both networking equipment and processing in the EOFC heuristic of Scenario 1 is higher than the EECC model when the percentage of traffic increase is more than 60%. This is because of the increase in the number of PSs utilized in the EOFC heuristic of Scenario 1 due to the limited link capacity at the access network, hence increasing the energy consumption of processing of the EOFC heuristic. Note that, the locations of the PSs are at both OLT and ONUs when the percentage increase in patients is more than 60%. The percentage increases in processing energy in the EOFC heuristic compared to the EECC model are as shown in Table 8.

Figure 12 also shows that the energy consumption of networking equipment of the EOFC heuristic under Scenario 2 is slightly higher than in Scenario 1. This is due to the additional energy consumed by the Ethernet switches at the access layer and the increasing utilization time of the networking equipment to send the analyzed health data storage traffic to the cloud storage in Scenario 2, hence high energy is obtained for EOFC heuristic of Scenario 2. Table 8 also shows that the energy saving of networking equipment of the EOFC heuristic of both scenarios compared to the EECC model decreases as the traffic increases. This is because the increase in energy consumption of networking equipment of the EOFC heuristic of both scenarios is higher than the EECC model. For instance, in Scenario 1, more networking equipment is utilized to place the PSs (OLT and ONU) compared to the EECC model. Meanwhile, for Scenario 2, the utilization time of the networking equipment to perform storage tasks in the EOFC heuristic is higher than in the EECC model. This is also due to the increasing energy consumption of the BS and ONU in both approaches to serve the raw health data traffic (EOFC heuristic of Scenario 1 and EECC model and EOFC heuristic of Scenario 2) which reduces the energy saving of networking equipment in the EOFC heuristic when compared to the EECC model.

Figure 12 also shows that the energy consumption for processing of the EOFC heuristic in Scenario 1 and Scenario 2 is slightly higher than in the EECC model for all percentages of traffic increase. Table 8 shows that the energy increase for processing of EOFC heuristic in Scenario 1 and Scenario 2 compared to the EECC model with 10% traffic increase reduced to 0.34% and 0.38%, respectively, compared to traffic in 2014/2015 which is 0.53%. This is due to the increasing utilization time of the PS in the EECC model to perform the feedback task since the increasing number of patients reduced the data rate allocated to each patient for the feedback transmission. However, the energy increase for processing for the EOFC heuristic compared to the EECC model under Scenario 2 increases with increasing traffic from 10% to 90%, as shown in Table 8. This is because the increase in total utilization time of the PSs in the EOFC heuristic under Scenario 2 is higher than in the EECC model for all percentages of traffic increase. Therefore, the processing energy increases for the EOFC heuristic compared to the EECC model under Scenario 2. Meanwhile, Table 8 also shows that the processing energy increase for EOFC heuristic under Scenario 1 is the same for traffic increase from 10% to 50%. This is mainly due to the same utilization time of the PSs to receive the raw health data and to transmit the analyzed health data for feedback and storage in the EOFC heuristic of Scenario 1 and the EECC model, regardless of the increase in the number of patients.

## VI. Performance Evaluation for Fall Monitoring Applications

This section presents the results and analysis of the EEFC model (i.e. FOA) for the fall monitoring application under two scenarios. The first scenario is with a limited number of patients per PS (Scenario 1) and the second scenario is with a limited number of PSs per candidate node (Scenario 2). Note that, as in the previous section, AMPL software with CPLEX 12.8 solver running on high-performance computing (HPC) cluster with a 12 core CPU and 64 GB RAM was used as a platform for solving the MILP model. Furthermore, the results of the EOFC heuristic running on a normal PC with 3.2 GHz CPU and 16 GB RAM are provided for a real-time implementation of the EEFC model. The same GPON architecture, as that shown in Figure 1, is used to evaluate the performance of the EEFC model and the EOFC heuristic in terms of the energy consumption of both the networking equipment and the processing. Note that, the PS used to perform the processing and the analysis of the video recording data is a 2.4-GHz Intel Core-Duo [59], that has maximum power consumption and idle power consumption of 85W and 10W, respectively.

### *1)* Limited number of patients per PS

In this section, the performance of both the EEFC model and the EOFC heuristic are evaluated for an increasing percentage of patients served in each PS. The conventional approach, the EECC model (i.e. CA), is used as the benchmark to evaluate the performance of both the EEFC model and the EOFC heuristic for the fall monitoring applications in terms of the energy consumption of both the networking equipment and the processing. Moreover, the optimization gaps between the EEFC model and the EOFC heuristic are presented in this section. Note that in this scenario, each candidate node can host more than 1 PS, hence $\phi_d$ is a variable. Table 9 shows the calculated input parameters for data rates and its transmission time for FOA and CA.

Table 9: Parameter inputs for FOA and CA for fall monitoring applications for Scenario 1

| Type of Data | Approach | Percentage of patients per PS | | | | |
|---|---|---|---|---|---|---|
| | | 20% | 40% | 60% | 80% | 100% |
| Data rate to transmit video to PS, $\delta a$ (kbps) | FOA | 15.456 | 15.792 | 16.128 | 16.800 | 17.136 |
| | CA | 15.456 | 15.792 | 16.128 | 16.464 | 17.136 |
| Transmission time to transmit video data to PS, $ta$ (s) | FOA | 217.39 | 212.77 | 208.33 | 200 | 196.08 |
| | CA | 217.39 | 212.77 | 208.33 | 204.08 | 196.08 |
| Data rate to transmit analyzed video to clinics, $\delta b$ (kbps) | FOA | 1.344 | 1.344 | 1.344 | 1.344 | 1.344 |
| | CA | 3.024 | 3.024 | 3.024 | 3.024 | 3.024 |
| Transmission time to transmit analyzed video to clinics, $tb$ (s) | FOA | 1.524 | 1.524 | 1.524 | 1.524 | 1.524 |
| | CA | 0.68 | 0.68 | 0.68 | 0.68 | 0.68 |
| Data rate to transmit analyzed video to cloud storage, $\delta c$ (kbps) | FOA | 1.674 | 1.674 | 1.674 | 1.674 | 1.674 |
| | CA | 38.571 | 38.571 | 38.571 | 38.571 | 38.571 |
| Transmission time to transmit analyzed video to cloud storage, $tc$ (s) | FOA | 1.223 | 1.223 | 1.223 | 1.223 | 1.223 |
| | CA | 0.053 | 0.053 | 0.053 | 0.053 | 0.053 |

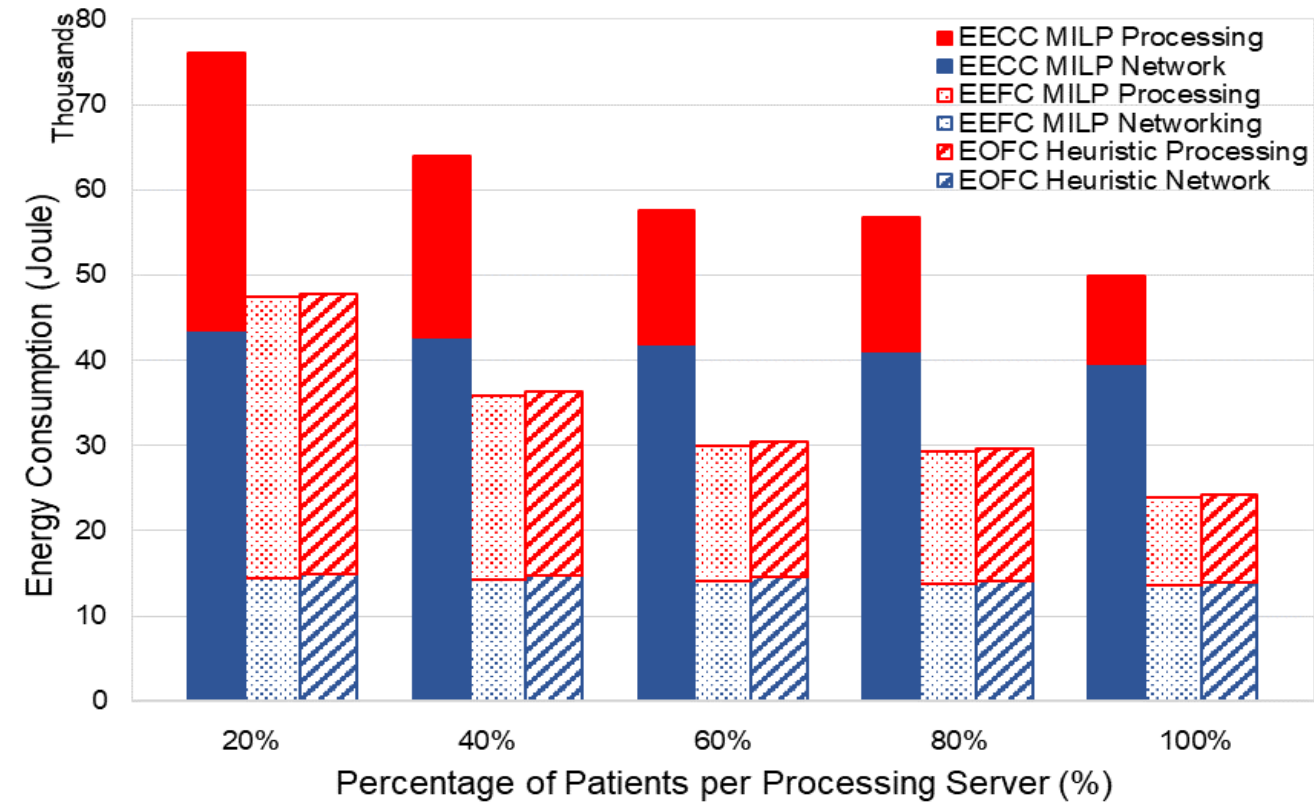

Figure 13: Energy consumption of networking equipment and processing for EECC model, EEFC model, and EOFC heuristic for different percentages of patients per PS

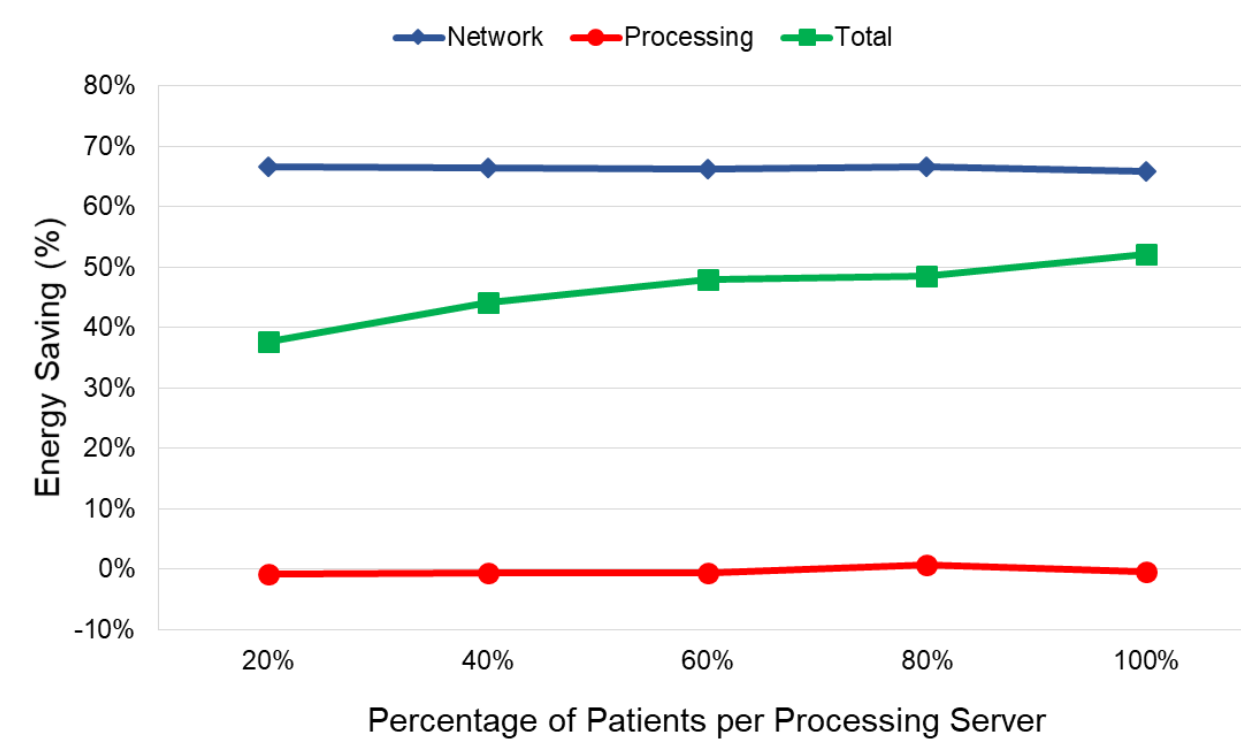

Figure 14: Percentage energy saving in EEFC model compared to EECC model for different percentages of patients per PS

Figure 13 shows the energy consumption of the networking equipment and the processing for the EECC model, EEFC model, and the EOFC heuristic, while Figure 14 shows the total energy saving, energy saving of the networking equipment, and the energy saving of the processing of the EEFC model as compared to those of the EECC model. The results are shown for an increasing percentage of patients that could be served at each PS. The results presented in Figure 13 revealed that the total energy consumption of the EEFC model is always smaller than that of the EECC model for all percentages of patients per PS. For instance, the total energy saving of the EEFC model compared to that of the EECC model is 38% when a single PS serves 20% of the total number of patients in the network, as shown in Figure 14. This saving is attributed to the fact that the location of the PSs in the EEFC model is the OLT, thereby reducing the amount of networking equipment utilized to transmit the raw health data traffic to the PS. Compared with the EECC model, the location of the PSs is in the cloud. Therefore, considerable energy is consumed in the metro and core layers to transmit the raw health data traffic to the PSs.

Figure 14 also shows that when a single PS serves 80% of the patients, the EEFC model saves 0.7% of the processing energy as compared to the EECC model. This saving is attributed to the low utilization time of the PS with the EEFC model to transmit the raw health data traffic to the PSs compared to the EECC model. Note that reducing the utilization time of the PSs reduces the energy consumption of the processing. Meanwhile, for the other percentages of patients served by a single PS, the amount of energy required for the processing in the EEFC model is slightly larger than that in the EECC model, as shown in Figure 13. The high energy consumption of processing in the EEFC model is attributed to the high utilization time of the PSs to send the analyzed health data feedback traffic and the analyzed health data storage traffic to the clinics and the cloud storage, respectively, compared to the EECC model, while the same amount of time is needed to transmit the raw health data traffic.

Figure 13 also shows that the total energy consumption of the EEFC model and the EECC model decreases when more patients can be served by a single PS. This is because allowing more patients to be served by a single PS reduces the number of utilized PSs, thereby reducing the processing energy consumption. Figure 14 shows that the total energy saving increases as the percentage of patients served by a single server increase. This is because allowing more patients to be served by a single PS reduces the available time to send the raw video recording to the PSs, which in turn reduces the energy consumed to keep the networking equipment and the PS in idle state.

Figure 14 also shows that the total energy consumption of the EOFC heuristic approached the total energy consumed by the EEFC model. Table 10 shows that the overall gap between the EEFC model and the EOFC heuristic for different percentages of patients per PS is less than 2%. This gap is mainly attributed to the higher number of base stations used by the EOFC heuristic compared to the EEFC model. The processing energy consumption of the EEFC model and the EOFC heuristic are equal as the same number of PSs is used.

Table 10: Optimization gap between the EEFC model and the EOFC heuristic for different percentages of patients per PS

| | Gap % | | | | |
|---|---|---|---|---|---|
| Percentage of patients per PS | 20% | 40% | 60% | 80% | 100% |
| Total energy | 0.98% | 1.26% | 1.47% | 1.45% | 1.74% |
| Network energy | 3.19% | 3.16% | 3.14% | 3.09% | 3.07% |
| Processing energy | 0% | 0% | 0% | 0% | 0% |

*2)* Limited number of PSs per candidate node

In this section, the performance evaluation of the EEFC model and EOFC heuristic for the increasing number of PSs per candidate node is presented. Moreover, the optimization performance gaps between the EEFC model and the EOFC heuristic are presented. Note that when the number of PSs per candidate node is limited to 1, we considered a single PS is connected directly to the ONU or the OLT and $\phi_d$ is set as parameter (i.e. $\phi_d = 1$). in contrast, when the number of PSs per candidate node is allowed to be more than 1 (i.e. $\phi_d$ is a variable), an Ethernet switch is used to connect the PSs to the ONU or the OLT. Also note that, in this scenario, each PSs can only serve 20% of the total patients. Table 11 shows the calculated input parameters of data rate and transmission time for FOA and CA.

Table 11: Input parameters for FOA and CA for scenario 2

| Type of Data | Number of PSs per candidate node | | | | |
|---|---|---|---|---|---|
| | 1 PS | 2 PSs | 3 PSs | 4 PSs | 5 PSs |
| Data rate to transmit video to PS, $\delta a$ (kbps) | 15.456 | 15.456 | 15.456 | 15.456 | 15.456 |
| Transmission time to transmit video to PS, $ta$ (s) | 217.39 | 217.39 | 217.39 | 217.39 | 217.39 |
| Data rate to transmit analyzed video to clinics, $\delta b$ (kbps) | 8.064 | 4.032 | 2.688 | 2.016 | 1.344 |
| Transmission time to transmit analyzed video to clinics, $tb$ (s) | 0.254 | 0.508 | 0.762 | 1.016 | 1.524 |
| Data rate to transmit analyzed video to cloud storage, $\delta c$ (kbps) | 8.370 | 4.185 | 2.79 | 2.092 | 1.674 |
| Transmission time to transmit analyzed video to cloud storage, $tc$ (s) | 0.245 | 0.489 | 0.734 | 0.979 | 1.223 |

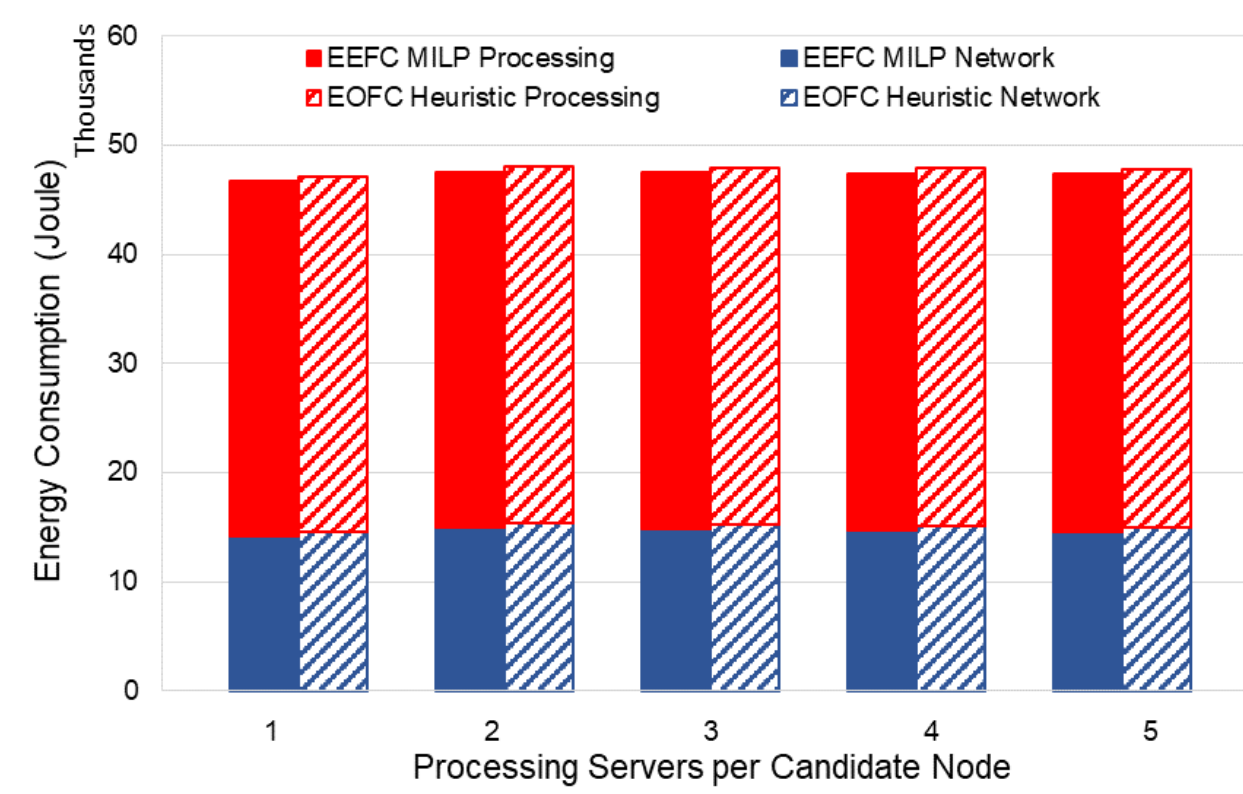

Figure 15: Energy consumption of networking equipment and processing for EEFC model and EOFC heuristic for different numbers of PSs per candidate node when 20% of patients can be served in a single PS

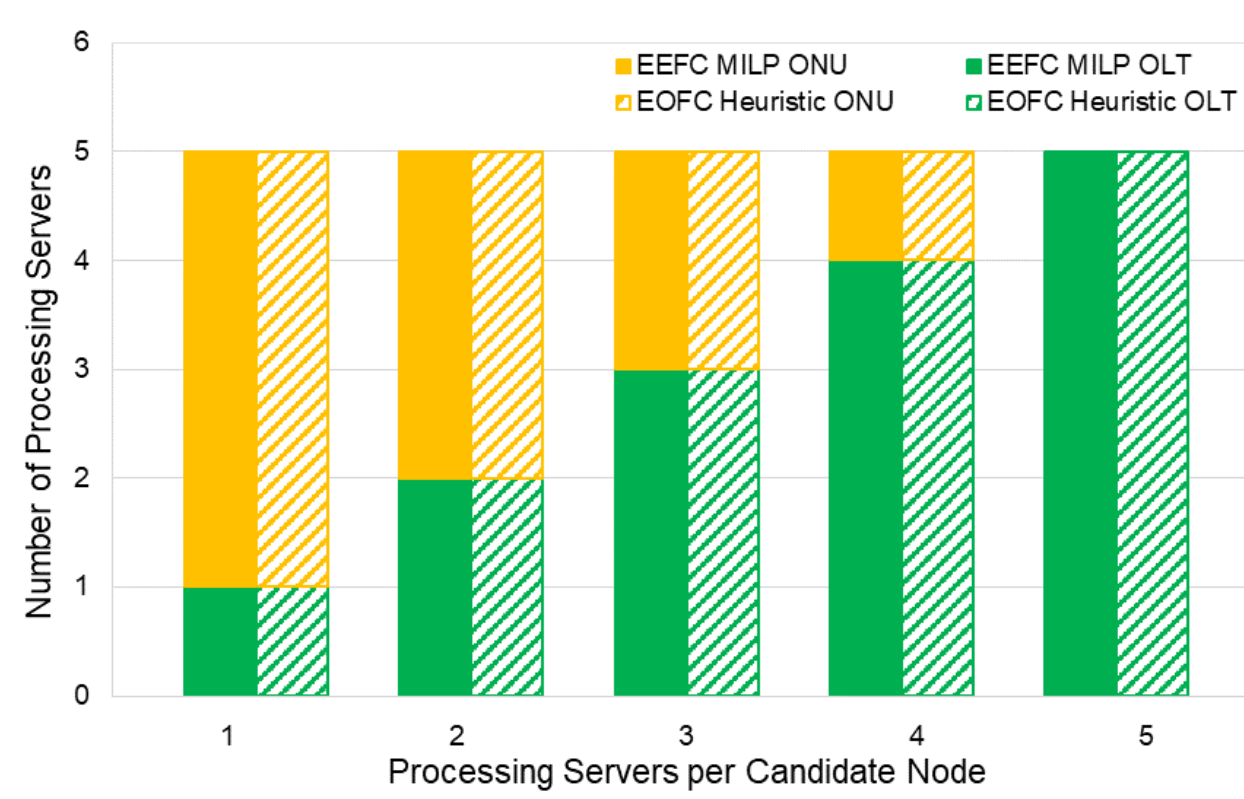

Figure 16: Optimal location of PSs for EEFC model and EOFC heuristic for different number of PSs per candidate node when 20% of patients can be served in a single PS

Table 12: Optimization gaps between the EEFC model and the EOFC heuristic for different numbers of PSs per candidate node

| | Gap (%) | | | | |
|---|---|---|---|---|---|
| Number of PSs per candidate node | 1 | 2 | 3 | 4 | 5 |
| Total energy | 0.99% | 0.97% | 0.97% | 0.97% | 0.98% |
| Network energy | 3.28% | 3.09% | 3.12% | 3.15% | 3.19% |
| Processing energy | 0% | 0% | 0% | 0% | 0% |

Figure 15 and Figure 16 show the total energy consumption of the networking equipment and the processing and the optimal location to place the PSs, respectively, for the EEFC model and the EOFC heuristic. Table 12 shows the optimization performance gaps between the EEFC MILP model and the EOFC heuristic. The results are shown for increasing number of PSs per candidate node. The results presented in Figure 15 revealed that the total energy consumption increases as the number of PSs per candidate node increased from 1 to 2. The increasing energy consumption is attributed to the utilization of the Ethernet switches dedicated to the health applications to connect multiple PSs to the ONU and the OLT.

Figure 15 also shows that the total energy consumption of both networking equipment and processing slightly decreases when the number of PSs per candidate node increased from 2 to 5. This is because limiting the number of PSs per candidate node required the placement of servers in multiple locations (i.e. OLT and ONUs) as opposed to the optimal location at the OLT when a node can accommodate multiple servers. This is shown in Figure 16 where more PSs are placed at the OLT while reducing the number of utilized ONUs to place the PSs when the number of PSs per candidate node increases. Note that the larger the number of nodes used to place the PS, the higher the energy consumption because of the increasing amount of networking equipment (i.e. Ethernet switches) used.

Furthermore, note that the data rate available per patient to transmit the analyzed health data feedback traffic and the analyzed health data storage traffic to the clinics and cloud storage, respectively, increases as fewer patients are served at a node. Hence, more time can be allocated to send the video signal to the PS. However, as the provisioned data rate to send the video signal is based on the number of PRBs, the same data rate and transmission time are used to send the video signal to the PS (i.e. irrespective of the number of PSs that can be placed at each candidate node), as shown in Table 11. This resulted in using the same number of base stations to serve all the patients. Note that the increasing energy due to the increasing amount of networking equipment used to host multiple PSs (i.e. Ethernet switches) with a small number of PSs at each candidate node dominated the reduction in energy. This is a result of the low utilization time of the network devices and the PSs used to transmit the analyzed health data feedback traffic and the analyzed health data storage traffic to the clinics and the cloud storage, respectively.

The results shown in Figure 15 also reveal that the total energy consumption (networking equipment and the processing) of the EOFC heuristic approached that of the EEFC model. The overall optimization gaps between the EOFC heuristic and the EEFC model for all numbers of PSs per candidate node is less than 1%, as shown in Table 12. This difference is only due to a large amount of networking equipment (i.e. base stations) used in the EOFC heuristic as compared to the EEFC model. However, the gap in energy consumption of the networking equipment between the EOFC heuristic and the EEFC model for all numbers of PSs per candidate node is less than 3.3%, as shown in Table 12.

## VII. Conclusions

This work has investigated the energy efficiency of an integrated healthcare approach that uses fog computing with the central cloud to serve low data rate and high data rate health monitoring applications. A PS is deployed at the access network to perform both processing and analysis of health data. The analyzed data is sent to the cloud for storage. A MILP model (EEFC) and a heuristic (EOFC) were developed to optimize the number and location of PSs at a GPON access network for energy-efficient fog computing. The results of the EEFC model for the low data rate health applications reveal that the optimal location for placing PSs is at the OLT as it is the nearest shared point to all the patients. The EEFC model achieved 36% total energy savings compared to the EECC model where the processing is performed at the central cloud. This saving is a result of reducing the traffic and the utilization time of the networking equipment. We also studied the impact of decreasing the idle power consumption of devices and increasing traffic volume on the performance of the EEFC model. The results revealed that the percentage network energy saving in the EEFC model compared to the EECC model decrease with decreasing percentage in idle power consumption of devices as idle power dominates the energy consumption of networking equipment and PS compared to its proportional load power. For high data rate applications, the results reveal that for scenario where the number of PSs can be hosted at each candidate node is not limited, a 38% total energy saving is achieved by the EEFC model as compared to the EECC model when 20% of the patients are served by a single PS. Increasing the number of patients served by a single PS reduced the total energy consumption of both the EEFC model and the EECC model because of the reduction in the number of activated PSs. The total energy saving in the EEFC model as compared to that in the EECC model increased to 52% when all the patients can be served by a single PS. Furthermore, the results show that increasing the number of PSs at each candidate node reduced the total energy consumption of the networking equipment and the processing. This reduction in energy is attributed to that fact that allowing more PSs at each candidate node reduced the total amount of networking equipment (i.e. Ethernet switches). The performance of the EOFC heuristic approaches that of the EEFC MILP model for low data rate and high data rate health applications.

## Acknowledgements

The authors would like to acknowledge funding from the Engineering and Physical Sciences Research Council (EPSRC), INTERNET (EP/H040536/1), STAR (EP/K016873/1) and TOWS (EP/S016570/1) and would like to acknowledge the Ministry of Education, Malaysia and Universiti Teknikal Malaysia Melaka (UTeM) for funding the PhD work of the first author. All data is provided in the results section of this paper.

## References

[1] A. Abayomi-Alli, A. J. Ikuomola, O. A. Aliyu, and O. Abayomi-Alli, "Development of a Mobile Remote Health Monitoring system – MRHMS," African Journal Computing & ICT, vol. 7, no. 4, pp. 15–22, 2014.

[2] P. L. and T. S. I. Azimi, A. Anzanpour, A. M. Rahmani, "Medical Warning System Based on Internet of Things Using Fog Computing," in 2016 International Workshop

on Big Data and Information Security (IWBIS), Jakarta, pp. 19–24, 2016.

[3] B. Xu, L. Xu, H. Cai, and L. Jiang, “Architecture of M-health monitoring system based on cloud computing for elderly homes application,” in 2014 Second International Conference on Enterprise Systems, pp. 45–50, 2014.

[4] H. Xia, I. Asif, and X. Zhao, “Cloud-ECG for real time ECG monitoring and analysis,” Computer Methods and Programs in Biomedicine, vol. 110, no. 3, pp. 253–259, 2012.

[5] N. K. and S. Z. K. Kaur, T. Dhand, “Container as a Service at the Edge : Trade - off between Energy Efficiency and Service Availability at Fog Nano Data Centers,” IEEE Wireless Communications, vol. 24, no. 3, pp. 48–56, 2017.

[6] T. N. Gia, M. Jiang, A. Rahmani, T. Westerlund, P. Liljeberg and H. Tenhunen, “Fog computing in healthcare Internet of Things: A case study on ECG feature extraction,” in 2015 IEEE International Conference on Computer and Information Technology; Ubiquitous Computing and Communications; Dependable, Autonomic and Secure Computing; Pervasive Intelligence and Computing, pp. 356–363, 2015.

[7] F. Jalali, K. Hinton, R. S. Ayre, T. Alpcan, and R. S. Tucker, “Fog Computing May Help to Save Energy in Cloud Computing,” IEEE Journal on Selected Areas in Communication, vol. 34, no. 5, pp. 1728–1739, 2016.

[8] L. F. Bittencourt, M. M. Lopes, I. Petri, and O. F. Rana, “Towards Virtual Machine Migration in Fog Computing,” in 2015 10th International Conference on P2P, Parallel, Grid, Cloud and Internet Computing (3PGCIC), pp. 1–8, 2015.

[9] S. K. Datta, C. Bonnet, and J. Haerri, “Fog Computing architecture to enable consumer centric Internet of Things services,” in 2015 International Symposium on Consumer Electronics (ISCE), Madrid, pp. 1-2, 2015.

[10] T. H. Luan, L. Gao, Z. Li, Y. Xiang, and L. Sun, “Fog Computing: Focusing on Mobile Users at the Edge,” *arXiv:1502.01815*, 2015.

[11] J. M. H. Elmirghani, T. Klein, K. Hinton, L. Nonde, A. Q. Lawey, T. E. H. El-Gorashi, M. O. I. Musa and X. Dong, “GreenTouch GreenMeter Core Network Energy-Efficiency Improvement Measures and Optimization,” IEEE/OSA Journal of Optical Communication and Networking, vol. 10, no. 2, pp. A250–A269, 2018.

[12] L. Nonde, T. E. H. El-gorashi, and J. M. H. Elmirghani, “Energy Efficient Virtual Network Embedding for Cloud Networks,” IEEE/OSA Jounal of Optical Communication and Networking, vol. 33, no. 9, pp. 1828–1849, 2015.

[13] A. Q. Lawey, T. E. H. El-Gorashi, and J. M. H. Elmirghani, “BitTorrent Content Distribution in Optical Networks,” Journal Lightwave Technology, vol. 32, no. 21, pp. 4209–4225, 2014.

[14] N. I. Osman, T. El-Gorashi, L. Krug, and J. M. H. Elmirghani, “Energy-efficient future high-definition TV,” Journal Lightwave Technology, vol. 32, no. 13, pp. 2364–2381, 2014.

[15] X. Dong, T. El-gorashi, and J. M. H. Elmirghani, “On the energy efficiency of physical topology design for IP over WDM networks,” IEEE/OSA Journal of Lightwave Technology, vol. 30, no. 12, pp. 1931–1942, 2012.

[16] X. Dong, T. El-Gorashi, and J. M. H. Elmirghani, “Green IP over WDM networks with data centers,” IEEE/OSA Journal of Lightwave Technology, vol. 29, no. 12, pp. 1861–1880, 2011.

[17] X. Dong, T. El-Gorashi, and J. M. H. Elmirghani, “IP Over WDM Networks Employing Renewable Energy Sources,” IEEE/OSA Journal of Lightwave Technology, vol. 29, no. 1, pp. 3–14, 2011.

[18] M. Musa, T. Elgorashi, and J. Elmirghani, “Energy efficient survivable IP-Over-WDM networks with network coding,” IEEE/OSA Journal Optical Communictions and Networking, vol. 9, no. 3, pp. 207–217, 2017.

[19] A. M. Al-Salim, A. Q. Lawey, T. E. H. El-Gorashi, and J. M. H. Elmirghani, “Energy Efficient Big Data Networks: Impact of Volume and Variety,” *IEEE* Transactions on Network and Service Management, vol. 15, no. 1, pp. 458–474, 2018.

[20] A. M. Rahmani, T.N. Gia, B. Negash, A.Anzanpour, I. Azimi, M. Jiang and P. Liljeberg, “Exploiting smart e-Health gateways at the edge of healthcare Internet-of-Things: A fog computing approach,” Future Generation Computer System, vol. 78, pp. 641–658, 2018.

[21] P. Verma and S. K. Sood, “Fog Assisted-IoT Enabled Patient Health Monitoring in Smart Homes,” IEEE Internet of Things Journal, vol. 5, pp. 1789-1796. 2017.

[22] M. O. I. Musa, T. E. H. El-gorashi, and J. M. H. Elmirghani, “Bounds on GreenTouch GreenMeter Network Energy Efficiency,” IEEE/OSA Journal of Lightwave Technology, vol. 36, no. 23, pp. 5395–5405, 2018.

[23] H. M. Mohammad Ali, T. E. H. El-Gorashi, A. Q. Lawey, and J. M. H. Elmirghani, “Future Energy Efficient Data Centers With Disaggregated Servers,” IEEE/OSA Journal of Lightwave Technology, vol. 35, no. 24, pp. 5361–5380, 2017.

[24] X. Dong, A. Lawey, T. E. H. El-Gorashi, and J. M. H. Elmirghani, “Energy-efficient core networks,” in 2012 16th International Conference on Optical Network Design and Modelling (ON*DM)*, Colchester , pp. 1-9, 2012.

[25] A. N. Al-quzweeni, A. Q. Lawey, T. E. H. El-Gorashi, and J. M. H. Elmirghani, “Optimized Energy Aware 5G Network Function Virtualization,” IEEE Access, vol. 7, pp. 44939–44958, 2019.

[26] Z. T. Al-Azez, A. Q. Lawey, T. E. H. El-Gorashi, and J. M. H. Elmirghani, “Energy Efficient IoT Virtualization Framework With Peer to Peer Networking and Processing,” *IEEE* Access, vol. 7, pp. 50697–50709, 2019.

[27] B. G. Bathula, M. Alresheedi, and J. M. H. Elmirghani, “Energy Efficient Architectures for Optical Networks,” Proceedings IEEE London Communications Symposium, London, pp. 5-8, 2009.

[28] B. G. Bathula and J. M. H. Elmirghani, “Energy efficient Optical Burst Switched (OBS) networks,” IEEE Globecom Workshops, Honolulu, pp. 1-6, 2009.

[29] T. E. H. El-Gorashi, X. Dong, and J. M. H. Elmirghani, "Green optical orthogonal frequency-division multiplexing networks," in IET Optoelectronics, vol. 8, no. 3, pp. 137–148, 2014.

[30] I. S. M. Isa, M. O. I. Musa, T. E. H. El-Gorashi, and J. M. H. Elmirghani, "Energy efficient and resilient infrastructure for fog computing health monitoring applications," in 2019 21st International Conference on Transparent Optical Networks (ICTON), Angers, France, pp. 1-5, 2019.

[31] I. S. M. Isa, T. E. H. El-Gorashi, M. O. I. Musa, and J. M. H. Elmirghani, "Resilient Energy Efficient Healthcare Monitoring Infrastructure with Server and Network Protection," Arvix 2020.

[32] A. Q. Lawey, T. E. H. El-Gorashi, and J. M. H. Elmirghani, "Distributed energy efficient clouds over core networks," Journal of Lightwave Technology, vol. 32, no. 7, pp. 1261–1281, 2014.

[33] A. M. Al-salim, T. E. H. El-gorashi, A. Q. Lawey, and J. M. H. Elmirghani, "Greening big data networks : velocity impact," IET Optoelectronics, vol. 12, no. 3, pp. 126–135, 2018.

[34] M. S. Hadi, A. Q. Lawey, T. E. H. El-gorashi, and J. M. H. Elmirghani, "Big data analytics for wireless and wired network design : A survey," Computer Networks vol. 132, pp. 180–199, 2018.

[35] M. S. Hadi, A. Q. Lawey, T. E. H. El-gorashi, and J. M. H. Elmirghani, "Patient-Centric Cellular Networks Optimization using Big Data Analytics," IEEE Access, vol. 7, pp. 49279–49296, 2019.

[36] M. Musa, T. Elgorashi, and J. Elmirghani, "Bounds for Energy Efficient Survivable IP over WDM Networks with Network Coding," IEEE/OSA Journal of Optical Communications and Networking, vol. 10, no. 5, pp. 471–481, 2018.

[37] I. S. M. Isa, M. O. I. Musa, T. E. H. El-gorashi, A. Q. Lawey, and J. M. H. Elmirghani, "Energy Efficiency of Fog Computing Health Monitoring Applications," in 2018 20th International Conference on Transparent Optical Networks (ICTON), Bucharest, pp. 1-5, 2018.

[38] I. Cale, A. Salihovic, and M. Ivekovic, "GPON (Gigabit Passive Optical Network)," in 29th International Conference on Information Technology Interfaces, pp. 679–684, 2007.

[39] A. D. Elgendi M, Eskofier B, Dokos S, "Revisiting QRS Detection Methodologies for Portable, Wearable, Battery-Operated, and Wireless ECG Systems," PLoS ONE 9(1) e84018, vol. 9, no. 1, 2014.

[40] N. Lowres, S. B. Freedman, R. Gallagher, A. Kirkness, D. Marshman, J. Orchard and L. Neubeck, "Identifying postoperative atrial fibrillation in cardiac surgical patients posthospital discharge, using iPhone ECG : a study protocol," BMJ Open, vol. 5:e006849, pp. 1–6, 2015.

[41] G. Mastorakis and D. Makris, "Fall detection system using Kinect's infrared sensor," Journal Real-Time Image Processing, vol. 9, no. 4, pp. 635–646, 2014.

[42] Age UK, "Later Life in the United Kingdom," 2015.

[43] Public Health England, "Public Health Profiles," Public Health England, 2015. [Online]. Available: http://healthierlives.phe.org.uk/. [Accessed: 20-Jun-2015].

[44] N. Hannent, "Ofcom UK Mobile Sitefinder," Ofcom UKMobile.[Online].Available:https://fusiontables.google.com/DataSource?docid=1N4nf1AmXFDk-lbh9Y54jh2FwyudbX3O8-aVlzwZJ#rows:id=1.

[45] Nokia, "LTE-M – Optimizing LTE for the Internet of Things White Paper," 2015.

[46] Eltex, "GPON Optical Line Terminal Data Sheet," 2015.

[47] "FTTC Exchanges," Sam Knows Ltd 2019, 2016. [Online].Available: https://availability.samknows.com/broadband/exchanges/bt/fttc.

[48] "Cardiovascular Disease Statistic 2017", British Heart Foundation, 2017.

[49] Offices for National Statistics, "Overview of the UK Population : November 2015," Office for National Statistics,2015.[Online].Available:http://www.ons.gov.uk/ons/rel/pop-estimate/overview-of-the-uk-population/index.html.

[50] Public HealthData Science Team, "Statistic of Patients with Heart Disease," Public Health England, 2018. [Online].Available:https://fingertips.phe.org.uk/profile-group/cardiovascular-disease-diabetes-kidney-disease/profile/cardiovascular.

[51] Cisco, "The Zettabyte Era : Trends and Analysis," 2017.

[52] R. Prieto, "Cisco Visual Networking Index Predicts Near-Tripling of IP Traffic by 2020," 2016.

[53] G. B. Moody and R. G. Mark, "The impact of the MIT-BIH Arrhythmia Database", in IEEE Engineering in Medicine and Biology Magazine, vol. 20, no. 3, pp. 45-50, 2001.

[54] A. L. Goldberger, L A Amaral, L Glass, J M. Haundorff, P C Ivanov, R G Mark, J E Mietus, G B Moody, C K Peng and H E Stanley, "PhysioBank, PhysioToolkit, and PhysioNet Components of a New Research Resource for Complex Physiologic Signals", Circulation, vol. 101, no. 23, pp. 215-220, 2000.

[55] A. L. Goldberger, Z. D. Goldberger, and A. Shvilkin, Goldberger's Clinical Electrocardiography. 2013.

[56] Stone, "STONEPC LITE," Stone Group, 2017. [Online].Available:https://www.stonegroup.co.uk/hardware/desktops/lite/.

[57] P. Kakria, N. K. Tripathi, and P. Kitipawang, "A real-time health monitoring system for remote cardiac patients using smartphone and wearable sensors", International Journal of Telemedicine and Applications, vol. 2015, pp. 1–11, 2015.

[58] Apple Inc., "Local and Push Notification Programming Guide," 2013.

[59] Apple Support, "iMac power consumption and thermal output,"2017.[Online].Available:https://support.apple.com/en-gb/HT201918.

[60] V. Valancius, N. Laoutaris, L. Massoulié, C. Diot, and P. Rodriguez, "Greening the internet with nano data centers," in 5th international conference on Emerging networking experiments and technologies CoNEXT 09, pp. 37-48, 2009.

[61] R. W. A. A. and R. S. T. A. Vishwanath, F. Jalali, K. Hinton, T. Alpcan, "Energy consumption comparison of interactive cloud-based and local applications," IEEE Journal on Selected Areas in Communications, vol. 33, no. 4, pp. 616–626, 2015.

[62] P. Mahadevan, P. Sharma, S. Banerjee, and P. Ranganathan, "A power benchmarking framework for network devices," International Conference on Research in Networking, pp. 795–808, 2009.
[63] Alcatel-Lucent, "Alcatel-Lucent 7368 ISAM ONT G-240G-A," 2014.
[64] J. Baliga, R. W. A. Ayre, K. Hinton, and R. S. Tucker, "Green Cloud Computing: Balancing Energy in Processing, Storage and Transport," in Proceedings of the IEEE, vol. 99, no. 1, pp. 149–167, 2011.
[65] C. Gray, R. Ayre, K. Hinton, and R. S. Tucker, "Power consumption of IoT access network technologies," in IEEE International Conference on Communication Workshop (ICCW), pp. 2818–2823, 2015.
[66] G. Auer, O. Blume, V. Giannini, I. Godor, and M. A. Imran, "Energy efficiency analysis of the reference systems, areas of improvements and target breakdown," EARTH Project Report Deliverable D2.3, pp. 1–68, 2012.
[67] NETGEAR, "NETGEAR 200 Series," 2015.

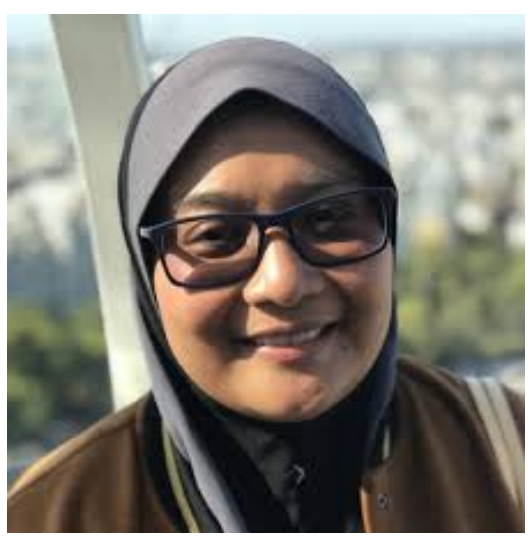

**Ida Syafiza M. Isa** obtained her PhD from the University of Leeds, UK in 2020 for work on energy efficient access networks design for healthcare applications. She has research interests in network architecture design, energy efficiency, network optimization, mixed integer linear programming and healthcare systems. She has published several papers in this area. She is currently a lecturer in Universiti Teknikal Malaysia Melaka (UTeM), Malaysia.

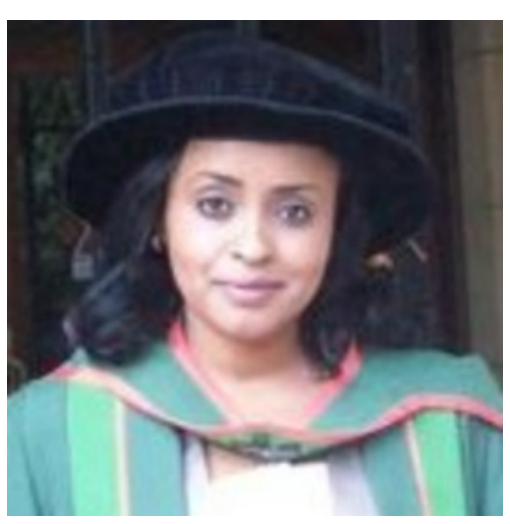

**Tasir E. H. EL-Gorashi** received the B.S. degree (first-class Hons.) in Electrical and Electronic Engineering from the University of Khartoum, Khartoum, Sudan, in 2004, the M.Sc. degree (with distinction) in Photonic and Communication Systems from the University of Wales, Swansea, UK, in 2005, and the PhD degree in Optical Networking from the University of Leeds, Leeds, UK, in 2010. She is currently a Lecturer in optical networks in the School of Electronic and Electrical Engineering, University of Leeds. Previously, she held a Postdoctoral Research post at the University of Leeds (2010– 2014), where she focused on the energy efficiency of optical networks investigating the use of renewable energy in core networks, green IP over WDM networks with datacenters, energy efficient physical topology design, energy efficiency of content distribution networks, distributed cloud computing, network virtualization and big data. In 2012, she was a BT Research Fellow, where she developed energy efficient hybrid wireless-optical broadband access networks and explored the dynamics of TV viewing behavior and program popularity. The energy efficiency techniques developed during her postdoctoral research contributed 3 out of the 8 carefully chosen core network energy efficiency improvement measures recommended by the GreenTouch consortium for every operator network worldwide. Her work led to several invited talks at GreenTouch, Bell Labs, Optical Network Design and Modelling conference, Optical Fiber Communications conference, International Conference on Computer Communications, EU Future Internet Assembly, IEEE Sustainable ICT Summit and IEEE 5G World Forum and collaboration with Nokia and Huawei.

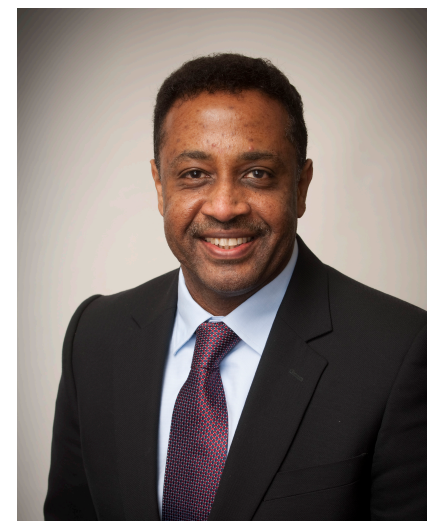

**Jaafar M. H. Elmirghani** is the Director of the Institute of Communication and Power Networks within the School of Electronic and Electrical Engineering, University of Leeds, UK. He joined Leeds in 2007 and prior to that (2000–2007) as chair in optical communications at the University of Wales Swansea he founded, developed and directed the Institute of Advanced Telecommunications and the Technium Digital (TD), a technology incubator/spin-off hub. He has provided outstanding leadership in a number of large research projects at the IAT and TD. He received the Ph.D. in the synchronization of optical systems and optical receiver design from the University of Huddersfield UK in 1994 and the DSc in Communication Systems and Networks from University of Leeds, UK, in 2012. He has co-authored Photonic switching Technology: Systems and Networks, (Wiley) and has published over 550 papers. He has research interests in optical systems and networks. Prof. Elmirghani is Fellow of the IET, Fellow of the Institute of Physics and Senior Member of IEEE. He was Chairman of IEEE Comsoc Transmission Access and Optical Systems technical committee and was Chairman of IEEE Comsoc Signal Processing and Communications Electronics technical committee, and an editor of IEEE Communications Magazine. He was founding Chair of the Advanced Signal Processing for Communication Symposium which started at IEEE GLOBECOM'99 and has continued since at every ICC and GLOBECOM. Prof. Elmirghani was also founding Chair of the first IEEE ICC/GLOBECOM optical symposium at GLOBECOM'00, the Future Photonic Network Technologies, Architectures and Protocols Symposium. He chaired this Symposium, which continues to date under different names. He was the founding chair of the first Green Track at ICC/GLOBECOM at GLOBECOM 2011, and is Chair of the IEEE Sustainable ICT Initiative, a pan IEEE Societies Initiative responsible for Green and Sustainable ICT activities across IEEE, 2012-present. He is and has been on the technical program committee of 41 IEEE ICC/GLOBECOM conferences between 1995 and 2021 including 19 times as Symposium Chair. He received the IEEE Communications Society Hal Sobol award, the IEEE Comsoc Chapter Achievement award for excellence in chapter activities (both in 2005), the University of Wales Swansea Outstanding Research Achievement Award, 2006, the IEEE Communications Society Signal Processing and Communication Electronics outstanding service award, 2009, a best paper award at IEEE ICC'2013, the IEEE Comsoc Transmission Access and Optical Systems outstanding Service award 2015 in recognition of "Leadership and Contributions to the Area of Green Communications", received the GreenTouch 1000x award in 2015 for "pioneering research contributions to the field of energy efficiency in telecommunications", the 2016 IET Optoelectronics Premium Award, shared with 6 GreenTouch innovators the 2016 Edison Award in the "Collective Disruption" Category for their work on the GreenMeter, an international competition, and received the IEEE Comsoc Transmission Access and Optical Systems outstanding Technical Achievement award 2020 in recognition of "Outstanding contributions to the energy efficiency of optical communication systems and networks", clear evidence of his seminal contributions to Green Communications which have a lasting impact on the environment (green) and society. He is currently an editor / associate editor of: IEEE Journal of Lightwave Technology, IEEE Communications Magazine, IET Optoelectronics, Journal of Optical Communications, and is Area Editor for IEEE Journal on Selected Areas in Communications (JSAC) Series on Machine Learning in Communication Networks (Area Editor). He was an editor of IEEE Communications Surveys and Tutorials and IEEE Journal on Selected Areas in Communications series on Green Communications and Networking. He was Co-Chair of the GreenTouch Wired, Core and Access Networks Working Group, an adviser to the Commonwealth Scholarship Commission, member of the Royal Society International Joint Projects Panel and member of the Engineering and Physical Sciences Research Council (EPSRC) College. He was Principal Investigator (PI) of the £6m EPSRC INTelligent Energy awaRe NETworks (INTERNET) Programme Grant, 2010-2016 and is currently PI of the £6.6m EPSRC Terabit Bidirectional Multi-user Optical Wireless System (TOWS) for 6G LiFi Programme Grant, 2019-2024. He has been awarded in excess of £30 million in grants to date from EPSRC, the EU and industry and has held prestigious fellowships funded by the Royal Society and by BT. He was an IEEE Comsoc Distinguished Lecturer 2013-2016.